\pdfoutput=1
\documentclass[cits]{JINST}

\usepackage{times}
\usepackage{lineno}
\usepackage{amsmath, amsthm, amssymb}
\usepackage{mathptmx}
\usepackage{mdwlist}
\usepackage{textcomp}
\usepackage{graphicx,subfigure}
\usepackage{fontenc}

\title{Measuring directionality in double-beta decay and neutrino interactions with kiloton-scale scintillation detectors}

\author{C.~Aberle$^a$, A.~Elagin$^b$, H.J.~Frisch$^b$, M.~Wetstein$^b$, and L.~Winslow$^a$\setcounter{footnote}{0}\thanks{corresponding author}\\
\llap{$^a$}University of California, Los Angeles, Los Angeles, CA 90095, USA\\
\llap{$^b$}University of Chicago, Chicago, IL 60637, USA\\
  E-mail: \email{lwinslow@physics.ucla.edu}}

\abstract{Large liquid-scintillator-based detectors have proven to be
exceptionally effective for low energy neutrino measurements due to
their good energy resolution and scalability to large volumes. The
addition of directional information using Cherenkov light and fast
timing would enhance the scientific reach of these detectors,
especially for searches for neutrino-less double-beta decay. In this
paper, we develop a technique for extracting particle direction using
the difference in arrival times for Cherenkov and scintillation light,
and evaluate several detector advances in timing, photodetector
spectral response, and scintillator emission spectra that could be
used to make direction reconstruction a reality in a kiloton-scale
detector.}


\begin{document}

\section{Introduction}
Liquid scintillator-based detectors are responsible for several of the
critical measurements that have determined our present understanding
of neutrino masses and mixings. These measurements include KamLAND's
measurement of reactor anti-neutrino oscillation at a distance of
$\sim$200~km\cite{kam2013}, Borexino's measurement of $^{7}$Be solar
neutrino oscillation\cite{borexino}, and most recently the short
baseline reactor anti-neutrino experiments that measured oscillations
due to $\theta_{13}$ at a distance of 1~km: Daya Bay\cite{dbtwo},
Double Chooz\cite{dctwo, dchydrogen}, and RENO\cite{reno}.
Scintillator-based neutrino detectors will continue to be important for the
next set of neutrino measurements, from the determination of the
neutrino mass hierarchy\cite{juno,reno50} to elastic scattering
measurements\cite{isodarscatt} and sterile neutrino
searches\cite{isodar,nist}, and for non-proliferation
applications\cite{nucifer, songs}.

The scalability of these detectors to large volumes also makes them
highly competitive for neutrino-less double-beta ($0\nu\beta\beta$)
decay searches in which the final state consists of a pair of
electrons with energies in the $\sim$1-2~MeV range.  The observation of this
rare decay would prove that the neutrino is a Majorana particle, which
would have profound consequences to our understanding of the generation of
mass and may provide a possible explanation of the matter-antimatter
asymmetry in the universe\cite{leptogenesis}.  Currently one of the best limits for the
$0\nu\beta\beta$ half-life comes from the scintillating detector
KamLAND-Zen\cite{KZ0nu}.

The advantage of liquid scintillators for measurements in the
$\sim$1~MeV range is their scalability from 1~ton to 1~kiloton while
providing energy resolutions of $\sigma(E)=\sim$5 \% $/\sqrt{E(MeV)}$\cite{kam2013,borexino}. This is roughly a factor of
two better than for water Cherenkov detectors, the other developed
technology for neutrino detectors that can be economically scaled to these large
masses. However, this energy resolution is much poorer than other technologies being used for $0\nu\beta\beta$ searches: Ge detectors\cite{gerda2013}, Te bolometers\cite{Alessandria:2011rc}, tracking detectors\cite{SuperNEMO}, liquid Xe time projection chambers (TPCs)\cite{EXO2012} and high pressure gaseous Xe TPCs\cite{NEXTsipm}.  

Scintillation light is isotropic. At these low energies, it does not contain sufficient information to reconstruct the track of the outgoing particles, although at higher energies it may\cite{john}.  Cherenkov light is also
produced for electrons above threshold. Most is absorbed and re-emitted as part of the
scintillation processes; however some fraction retains its
directional information. If this directional Cherenkov light can be
isolated from the copious isotropic scintillation light, it may be
possible to reconstruct the direction of the primary particle. The addition of directionality is a powerful tool
for background rejection, especially for $0\nu\beta\beta$ searches. In high pressure TPCs, reduction factors on the order of $\sim10^4$ have been achieved\cite{Gotthard}. It is also possible to look for new physics in the angular correlation of the emerging electrons\cite{newphysics0nuBB}, as has been proposed for the tracking-based detectors\cite{SuperNEMO,NEXTsipm}. The addition of a directional signal would make large-scale liquid scintillator detectors more competitive for the next generation of $0\nu\beta\beta$ searches. 

This is the first in a series of papers exploring directionality in large-scale liquid scintillator detectors. In this paper, we develop a technique for
separating the Cherenkov and scintillation light using the photon arrival
times and evaluate several detector advances in timing, photodetector 
spectral response, and scintillator emission spectra that would allow
the realization of direction reconstruction in kilo-ton scale
scintillating neutrino detectors. This is different from the direction reconstruction described for high-energy neutrino interactions\cite{john} or that for neutrons from inverse beta decay\cite{chooz,dcDirection}.  We then use these results as input into a traditional direction reconstruction developed for water Cherenkov detectors. Since the reconstruction of the direction of $\sim$1~MeV particles has not been achieved before, we start these studies with the 
simple case of a single particle at the center of the detector. We also start with an easier test case of a 5~MeV electron like that from a $^{8}$B solar neutrino interaction. With the higher photon statistics at this energy, we verify the technique and the different detector parameters that affect it. We then study the technique for two lower energies, 1.4 and 2.1 MeV, that are more relevant to $0\nu\beta\beta$.  

\section{Liquid scintillator detectors}
Liquid scintillators are `cocktails' of aromatic hydrocarbons. When
charged particles move through a scintillator, the molecules are
excited, predominantly via the non-localized electrons in the
$\pi$-bonds of the phenyl groups\cite{birks_book}. Vibrational
and rotational modes of the molecules are turned into heat within
picoseconds through collisions with other molecules.  Within $\sim$10
picoseconds, the $\pi$-electrons de-excite to the first excited state
from higher levels through radiationless transitions. The first
excited state can de-excite through photon emission. There are two
characteristic times for this de-excitation, depending if the singlet
state or the triplet state was excited.  The singlet state will
de-excite within nanoseconds while the triplet state de-excites on the
order of 10's or 100's of nanoseconds. These two processes are
fluorescence and phosphorescence respectively. The exact time
constants for these processes are determined by the composition of the
scintillator.

The absorption and emission spectra overlap at some level
in all molecules. Consequently, if there is only one type of molecule in
the scintillator cocktail the light output is reduced due to
inefficiencies in the energy transfer through multiple absorption and
re-emission processes. Aromatic solutes or fluorophores are added to the
primary solvent to shift the wavelengths of the photons to higher values 
where the scintillator is more transparent. This
wavelength-shifting is also used to match the quantum efficiency as a
function of wavelength for the photodetectors being used. One typical
scintillator mixture uses pseudocumene(1,2,4-trimethylbenzene) as the solvent with 1-5~g/l of
PPO (2,5-diphenyloxazole) as the fluorophore. This mixture has a peak emission at about 400~nm
where bialkali photomultiplier tubes (PMTs) are most sensitive and the
pseudocumene is relatively transparent.

A good liquid scintillator will produce $\sim$10,000 photons
isotropically per MeV of deposited energy. Although less abundant,
Cherenkov light will be produced as well if a particle is moving
faster than the speed of light in the medium.  This light is emitted
in a cone centered on the direction of the particle trajectory, and
with a continuous spectrum weighted toward shorter wavelengths but
extending well into the red. The spectrum is described
by\cite{Cherenkov34}:
\begin{equation}
\label{eqCherenkov}
\frac{d^2N}{d\lambda dx} = \frac{2 \pi \alpha Z^2}{\lambda^2} \left [ 1 - \frac{1}{\beta^2 n(\lambda)^2} \right ]
\end{equation}
where $N$ is the number of Cherenkov photons, $\lambda$ is the wavelength of the photon, x is the distance travelled by the particle, $Z$ is the charge of the particle, $\alpha$ is the fine structure constant, $n(\lambda)$ is the wavelength-dependent index of refraction and
$\beta$ is the velocity of the particle. 

The Cherenkov light produced at wavelengths shorter than the
absorption cutoff of the scintillator will be absorbed and re-emitted
as isotropic light, but wavelengths longer than this cutoff will
propagate across the detector, retaining their directional
information. The absorption cutoff for the example scintillator above is 370~nm. The index of refraction at $\lambda$=370~nm is n=1.466, which translates
into an effective Cherenkov threshold for electrons at 0.188~MeV. For a 5~MeV electron, this yields 685 Cherenkov photons per event when integrating from the cutoff wavelength at 370~nm to 550~nm,  the wavelength above which a small number of photons is detected due to the quantum efficiency of a standard bialkali photocathode. Lowering the energy to 1~MeV yields 82 Cherenkov photons between 370-550~nm.


All photons including these undisturbed Cherenkov
photons will have timing determined by the group 
velocity\cite{group_velocity_article,pdg_review_2012,tamm1939} in the liquid,
\begin{equation}
\label{eqGroup}
v_{g}(\lambda) = \frac{c_{vacuum}}{n(\lambda) - dn(\lambda)/d\textnormal{log}(\lambda)}.
\end{equation}
Photons at the scintillation cutoff of 370~nm have a velocity of 0.191 m/ns, while photons with wavelengths of 600 nm are appreciably faster, with a velocity of 0.203 m/ns. Since on average the undisturbed Cherenkov photons have longer wavelengths, they will arrive before the
scintillation light, which is slowed by both the scintillation
processes and the shorter wavelengths involved. Thus, with sufficient
timing resolution and sensitivity to longer wavelengths it should be
possible to separate the directional Cherenkov light and the isotropic
scintillation light, and then to reconstruct the direction of the
initial particle.

In neutrino-electron scattering events, a single electron emerges with a distribution of energies with a maximum energy related to the incoming neutrino's energy. In comparison, $0\nu\beta\beta$ events are more complicated, with two electrons emerging with a combined energy equal
to the Q-value of the particular isotope. The individual
electrons follow distributions of energies and angular correlations which depend on the underlying $0\nu\beta\beta$ decay mechanism\cite{SuperNEMO,newphysics0nuBB,bandv}. Figure 1 shows a simulated $^{116}$Cd $0\nu\beta\beta$ event in a model with light Majorana neutrino exchange, for which a probable case is the emission of two electrons with comparable energies at a large angle relative to each other. 

High Q-value candidates are preferred for $0\nu\beta\beta$  for two reasons. First, as the measured half-life is inversely proportional to the phase-space factor, the measured rate is expected to be higher from isotopes with higher Q-values. Second, the main background for these experiments come from the daughters of the $^{238}$U and $^{232}$Th decay chains. The Compton shoulder from gamma-rays is particularly problematic. The 2.6~MeV gamma-ray from $^{208}$Tl decay is the highest energy gamma from these decay chains and so isotopes above 2.6~MeV are preferred.  Experiments using isotopes with Q-values below this energy must compensate with improved background rejection techniques.

Most of the high Q-value candidates\cite{tabledbb}
have been considered as a dopant for a liquid scintillator:
$^{150}$Nd (Q=3.367~MeV)\cite{minfang,nd1}, $^{96}$Zr (Q=3.350~MeV)\cite{zr1},
$^{100}$Mo (Q=3.034~MeV)\cite{mo1}, $^{82}$Se (Q=2.995~MeV)\cite{qdot},
$^{116}$Cd (Q=2.81~MeV)\cite{qdot, cd1}, $^{130}$Te (Q=2.533~MeV)\cite{qdot, biller},
$^{136}$Xe (Q=2.479~MeV)\cite{KZ0nu} and $^{124}$Sn (Q=2.29~MeV)\cite{sn1}.
Xenon gas readily dissolves into liquid scintillator. For the other isotopes,
a suitable organometallic compound needs to be found that produces a stable
scintillator with a long attenuation length in the wavelength region of
interest. Recently nanocrystals formed by candidate isotopes have been explored as an alternative to doping by single atoms\cite{qdot,qdot2}.

\begin{figure}
        \begin{center}
        \includegraphics[scale=0.4]{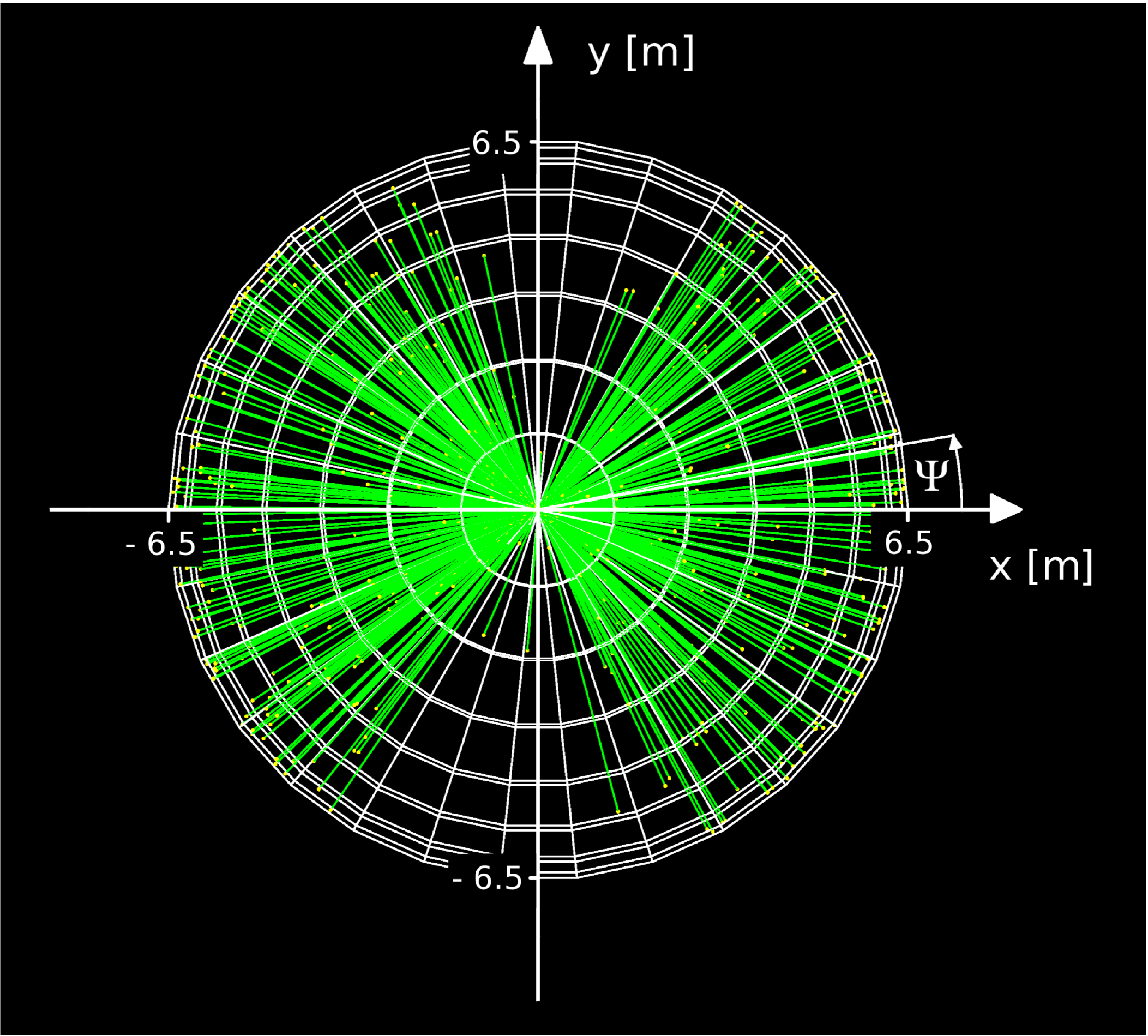}
        \caption[]{The detector geometry and coordinate system.
        The radial rays (green lines) are photons emitted by two back-to-back electrons with 1.4~MeV each
        (equally divided energy of $^{116}$Cd~$0\nu\beta\beta$ decay). The electrons originate at
        the center of the sphere with initial directions along the x
        and -x-axis. Only Cherenkov photons are drawn to illustrate the
        directionality of the event. \label{detector_view}}
        \end{center}
\end{figure}

\section{Geant4 simulation}
\label{sim_section}
In order to study the effects relevant to directional reconstruction
in liquid scintillators, a Geant4\cite{geant4one,geant4two} simulation
has been constructed. The simulation uses Geant4~version 4.9.6 with the default liquid scintillator
optical model, in which optical photons are
assigned the group velocity in the wavelength region of normal
dispersion.

The detector geometry is a sphere of 6.5~m radius filled with
scintillator. Figure \ref{detector_view} shows the geometry and the
Cherenkov light from an example $^{116}$Cd $0\nu\beta\beta$ event. The
default scintillator properties have been chosen to match a KamLAND-like
scintillator\cite{kamland2003}: 80\% n-dodecane, 20\% pseudocumene and 1.52~g/l PPO. The
scintillator properties implemented in the simulation include the
atomic composition and density ($\rho$ = 0.78~g/ml), the
wavelength-dependent attenuation length\cite{tajimaMaster} and
refractive index\cite{OlegThesis}, the scintillation emission
spectrum\cite{tajimaMaster}, emission rise time ($\tau_r$ = 1.0~ns)
and emission decay time constants ($\tau_{d1}$ = 6.9~ns and
$\tau_{d2}$ = 8.8~ns with relative weights of 0.87 and 
0.13)\cite{tajimaThesis}, scintillator light yield (9030 photons/MeV),
and the Birks constant ($kB$ $\approx$ 0.1~mm/MeV)\cite{ChrisThesis}.  This is a standard scintillator. The attenuation length at 400~nm, the position of the peak standard bialkali photocathode efficiency, is 25~m. The attenuation length drops precipitously between 370~nm and 360~nm from 6.5~m to 0.65~m. We use this drop to define the cutoff wavelength at 370~nm. Variations from the baseline KamLAND case are discussed below. 

\begin{figure}
        \begin{center}
        \includegraphics[scale=0.4]{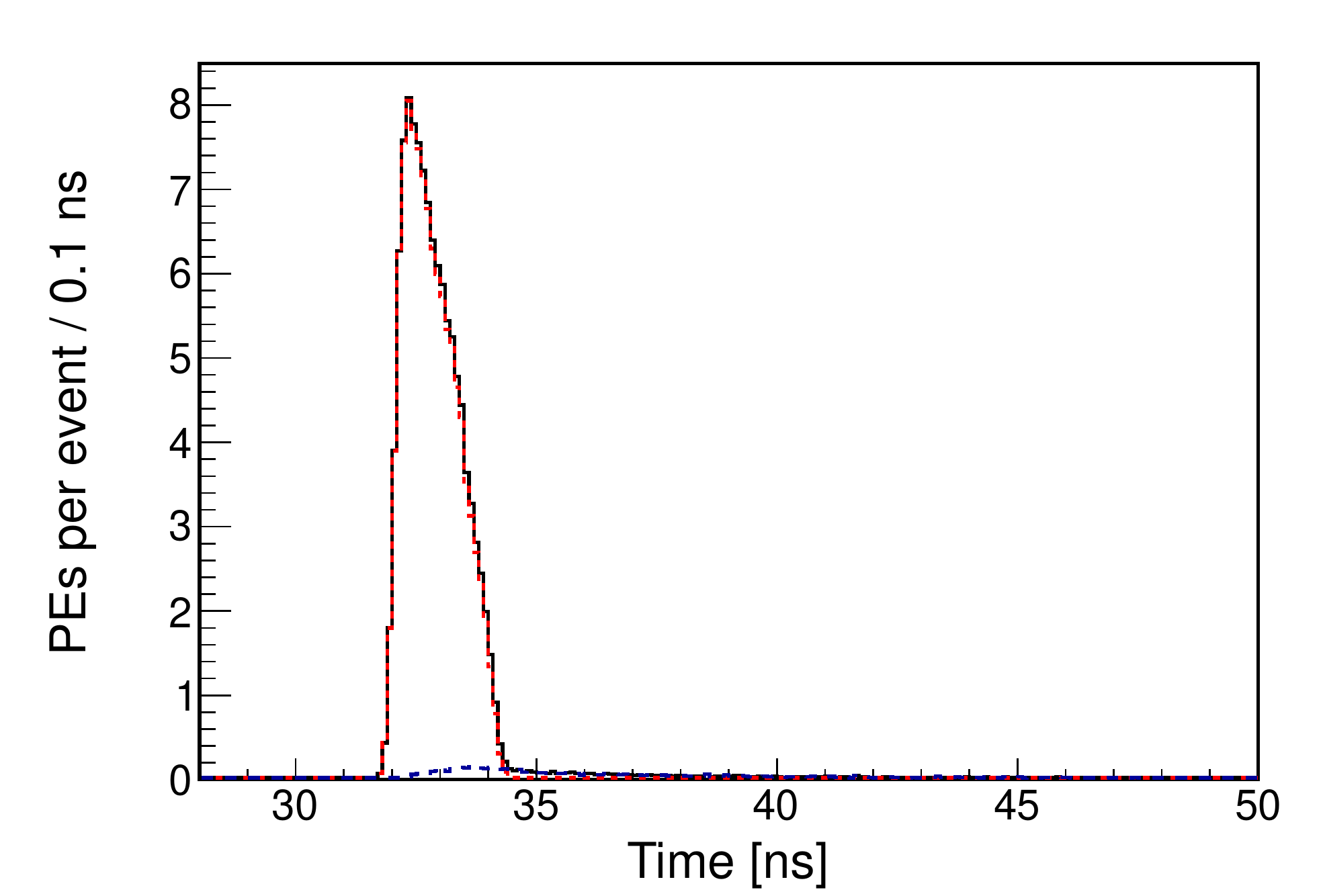}
        \caption[]{Cherenkov photoelectron (PE) arrival times after application
        of the transit-time spread (TTS) for the simulation
        of 1000 electrons (5~MeV). The default simulation is shown with optical scattering turned on: all photons (black solid), un-scattered photons (red dotted), and scattered (blue dashed). Scattering causes a very small tail at longer times as expected. \label{scattplot}}
        \end{center}
\end{figure}

Re-emission of absorbed photons in the scintillator
bulk volume and optical scattering, specifically Rayleigh scattering, have not yet been included by default. A test simulation shows that the effect of optical scattering is negligible. As shown in figure~\ref{scattplot}, scattering causes a small tail at longer times. The reason is that the cutoff is very steep below 360~nm and almost no photons reach the sphere, so optical scattering makes no difference for short wavelengths. Above about 395~nm, the attenuation length is greater than 20~m so both scattering and absorption are not very likely and scattering is negligible. The intermediate region is rather small. A similar argument holds for re-emission. Scattering length measurements and discussions can be found in Ref.~\cite{Wurm:2010ad}. 

The inner sphere surface is used as the photodetector. It is treated
as fully absorbing (no reflections), with a photodetector coverage of
100\%. As in the case of optical scattering, reflections at the sphere are a small effect that would create a small tail at longer times. Two important photodetector properties have been varied: 1)
the transit-time spread (TTS, default $\sigma$ = 0.1~ns) and 2) the
wavelength-dependent quantum efficiency (QE) for photoelectron
production. The default is the QE of a bialkali photocathode (Hamamatsu
R7081 PMT)\cite{Hamamatsu_R7081}. The QE values as a function of wavelength come from the Double Chooz\cite{dctwo}
Monte Carlo simulation. We note that the KamLAND 17-inch PMTs use the
same photocathode type with similar quantum efficiency. We are neglecting any threshold effects in the photodetector readout electronics.

\begin{figure}
\begin{center}
        \subfigure[ ~Default
        simulation.]{\includegraphics[scale=0.295]{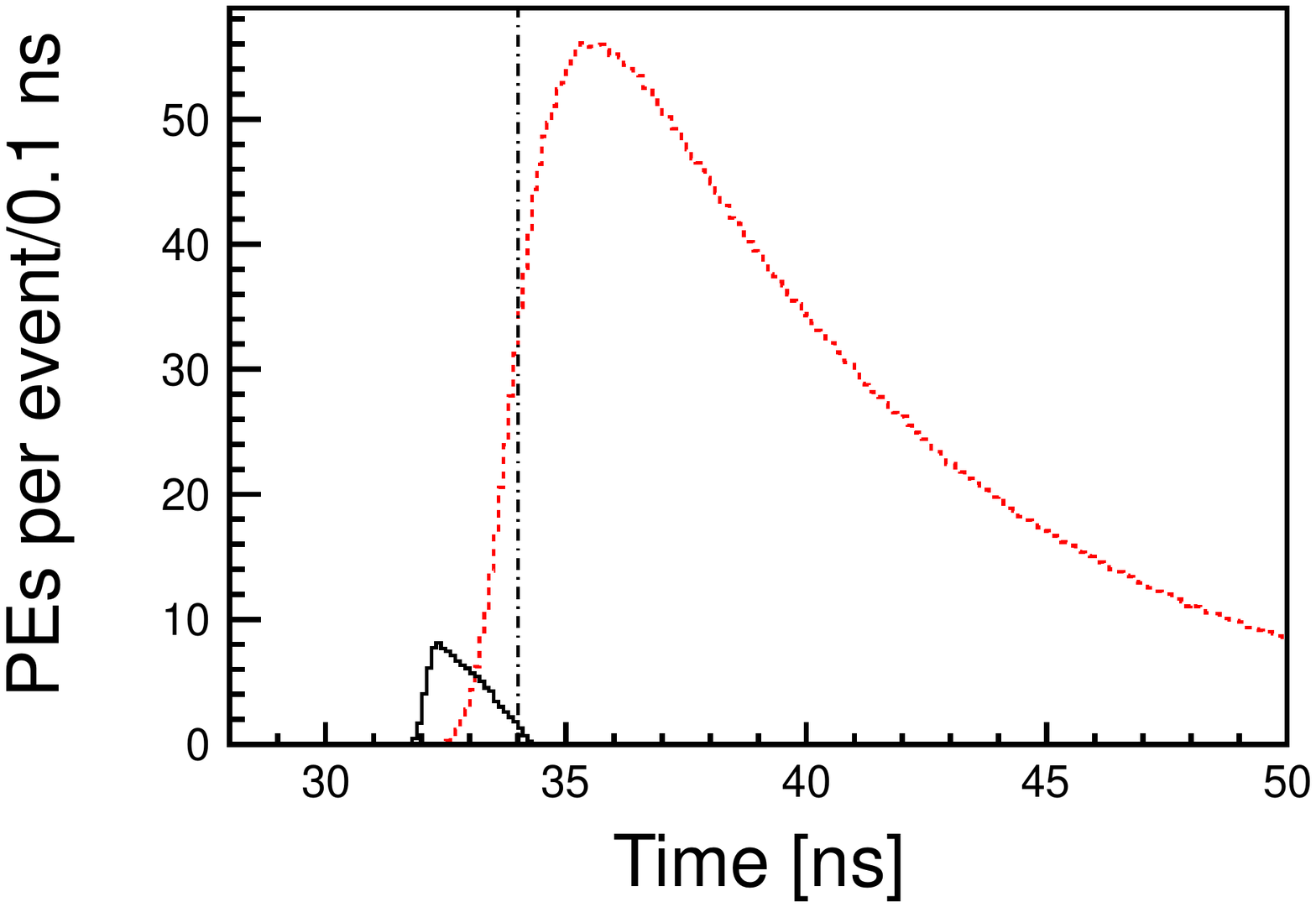}}
        \subfigure[ ~Increased TTS
        (1.28~ns).]{\includegraphics[scale=0.295]{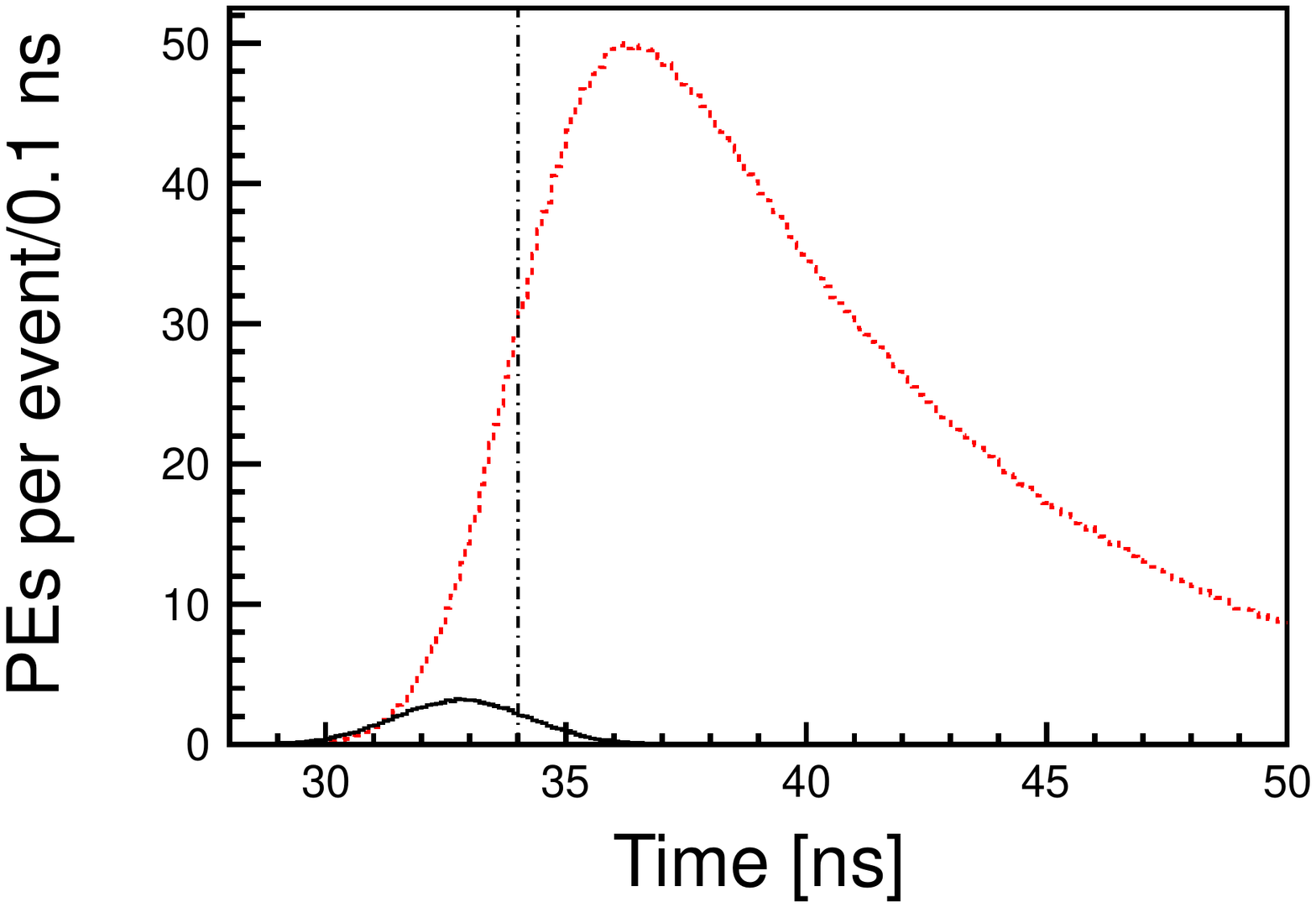}}
        \subfigure[ ~Red-sensitive
        photocathode.]{\includegraphics[scale=0.295]{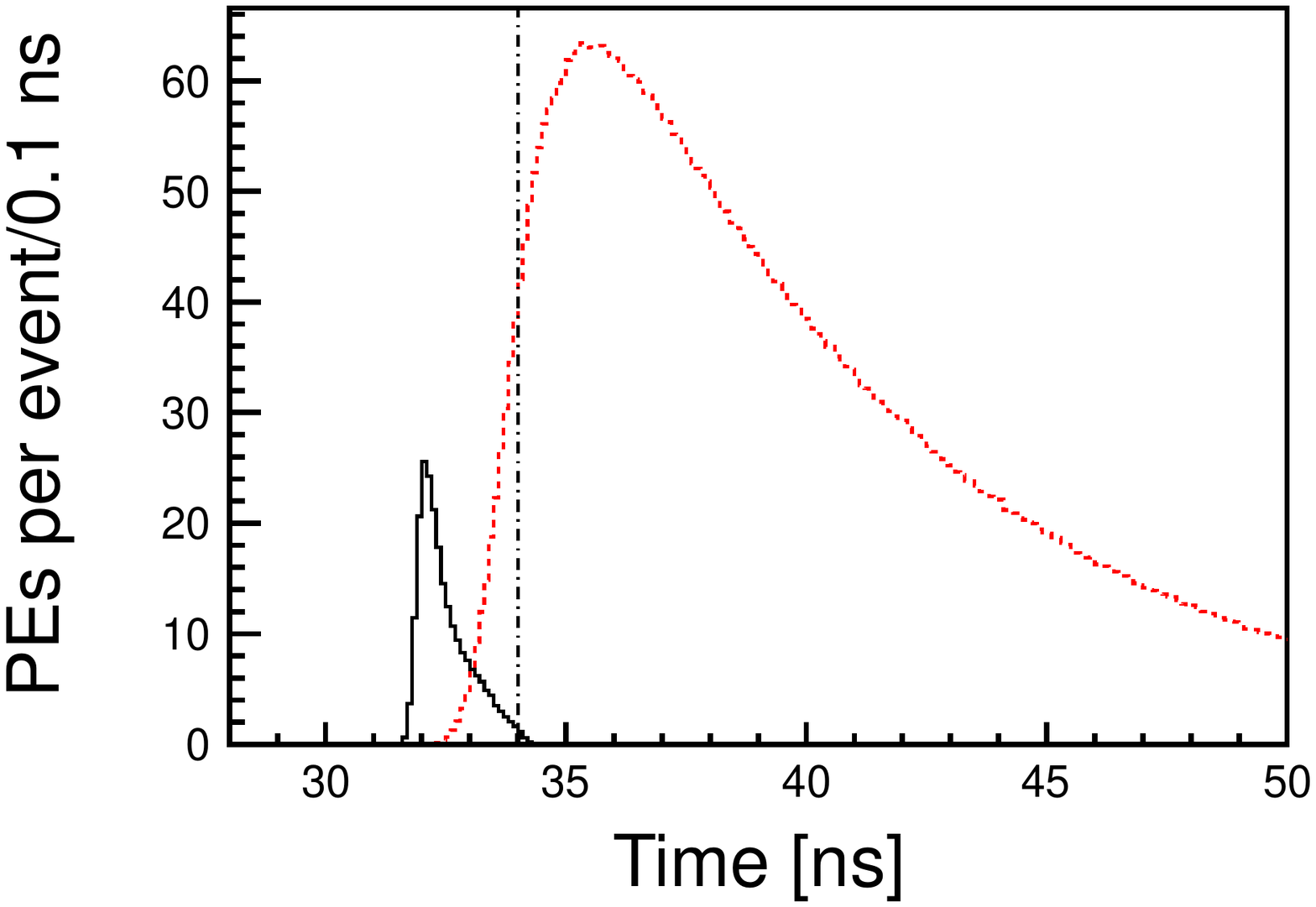}}
        \caption[]{Photoelectron (PE) arrival times after application
        of the transit-time spread (TTS) for the simulation
        of 1000 electrons (5~MeV) with different values of the TTS and
        wavelength response. PEs from
        Cherenkov light (black, solid line) and scintillation light
        (red, dotted line) are
        compared. The dash-dotted vertical line illustrates a time cut at
        34.0~ns. (a) Default simulation: bialkali photocathode and TTS =
        0.1~ns ($\sigma$). After the 34.0~ns time cut, 171~PEs
        from scintillation and 108~PEs from Cherenkov light are detected. (b)
        Default simulation settings except for TTS = 1.28~ns (KamLAND
        17-inch PMTs). After the 34.0~ns time cut, 349~PEs from
        scintillation and 88~PEs from Cherenkov light are detected. (c) Default
        simulation settings except for a GaAsP photocathode. After the
        34.0~ns time cut, 226~PEs from scintillation and 229~EPS
        from Cherenkov light are detected. \label{time_plots_comparison}}
        \end{center}
\end{figure}


Four effects primarily contribute to the timing of the scintillator detector
system: the travel time of the particle, the time constants of the scintillation process, chromatic dispersion, and the timing of the photodetector.  First, the simulated travel time of a 5~MeV electron is 0.108$\pm$0.015~ns. This corresponds to an average path length of 3.1~cm and a final distance from the origin of 2.6~cm. The time until the electron drops below Cherenkov threshold is 0.106$\pm$0.015~ns. We note that due to scattering the final direction of the electron before it stops does not correspond to the initial direction; however the scattering angle is small while the majority of Cherenkov light is produced. The Cherenkov light thus encodes the direction of the primary electron. The scattering physics is handled by Geant4's ``Multiple Scattering" process which is valid down to 1~keV, where atomic shell structure becomes important\cite{geant4scatt}. In the energy range important for $0\nu\beta\beta$, a 1.4~MeV electron travels a total path length of 0.8~cm, has a distance from the origin of 0.6~cm in 0.030$\pm$0.004~ns  and takes 0.028$\pm$0.004~ns to drop below Cherenkov threshold. The scattering follows the same pattern. 

The scintillator-specific rise and decay times are the second effect that determines the timing in a scintillator detector. The first step in the scintillation process is the transfer of energy from the solvent to the solute. The time constant of this
energy transfer accounts for a rise time in scintillation light
emission. Past neutrino experiments were not highly sensitive to the
effect of the scintillation rise time, which is the reason why there
is a lack of accurate numbers. We assume a rise time of 1.0~ns; more
detailed studies are needed in the future. The two time constants used
to describe the falling edge of the scintillator emission time
distribution (quoted above) are values specific to the KamLAND
scintillator.

Chromatic dispersion is the third effect that determines the timing in a scintillator detector. Due to the wavelength-dependence of the refractive index, the speed of
light in the scintillator (see Equation (\ref{eqGroup})) increases
with increasing photon wavelengths for normal dispersion, with red
light traveling faster than blue light.  In order to study the time differences due to this chromatic dispersion, we used a simplified simulation of 5~MeV electrons at
the center of the sphere where we used instantaneous scintillation
emission with the quantum-efficiency applied, but not including a
transit-time spread. The higher energy 5~MeV electron provides larger photon statistics with which to evaluate the method, debug the simulation, and is of interest to neutrino-electron scattering experiments. The true hit time distributions of photoelectrons
were analyzed for scintillation light and Cherenkov light
separately. Photoelectrons coming from Cherenkov light are on average
created about 0.5~ns earlier than PEs from scintillation light. The
RMS values from PE time distributions for Cherenkov and scintillation
light are both about 0.5~ns. Note that these numbers include the
effect of the finite electron travel time.

The fourth effect determining the timing in a scintillator detector is the timing of the photodetectors. The measurement of the arrival times of single photoelectrons is
affected by the transit-time spread (TTS) of the photodetectors, a
number which can be different by orders of magnitude depending on the
detector type. The default TTS of 0.1~ns ($\sigma$) can be achieved with large area picosecond photodetectors
(LAPPDs)\cite{Adams:2013nva,RSI_paper,PSEC4_paper,anode_paper} and possibly hybrid photodetectors
(HPDs)\cite{hpdThesis}; even significantly lower TTS numbers are
realistic with the LAPPD\cite{RSI_paper,PSEC4_paper,anode_paper}.

In sections \ref{detector_timing_sec} to
\ref{scintillator_emission_sec}, we study the
photoelectron timing for different detector configurations at 5~MeV. We focus
on the idea of increasing the discrimination between Cherenkov and
scintillation light by using improved detector timing. The primary
quantities provided by the Geant4~simulation are the photoelectron hit
positions and the detection times after the TTS resolution has been
applied. In section~\ref{reconstruction_sec} these quantities are then
used for event reconstruction. With the successful reconstruction at 5~MeV, we then lower the energy of the simulated electrons and show that it is possible to reconstruct electrons in the range interesting for $0\nu\beta\beta$.

\section{Detector timing}
\label{detector_timing_sec}

We first discuss results for the default simulation settings described
in the previous section. Figure \ref{time_plots_comparison} (a) shows
the TTS-smeared photoelectron (PE) detection times for 1000 simulated
electrons with 5~MeV energy in the center of the detector, with initial
momentum directions coinciding with the x-axis. The photoelectrons
induced by Cherenkov light arrive earlier, as expected due to the
instantaneous emission and the higher average photon speed compared to
scintillation light. There is, however, significant overlap of the two
arrival time distributions.

In order to compare simulations with different parameters, a fixed
time cut of $t\leq$ 34.0~ns is applied using the truth information to
isolate the Cherenkov light in this early time window, as shown in figure \ref{time_plots_comparison} (a). For the default
simulation case, the average number of PEs per event coming from
Cherenkov light in the early time window (108) is 98\% of the total
average number of PEs from Cherenkov light (110). For scintillation
light, the average number of PEs (171) is only 3.1\% of the average total
scintillation-induced PEs (5445). This demonstrates the effectiveness of
a time cut to separate Cherenkov light from scintillation light. 

The ratio of Cherenkov-induced to scintillation-induced photoelectrons
in the early time window ($R_{C/S}$) is a useful figure-of-merit when
comparing different simulation settings, since a higher ratio means
more directional information per PE. For the default simulation
settings $R_{C/S}=0.63$.

\begin{figure}
        \begin{center}
        \includegraphics[scale=0.40]{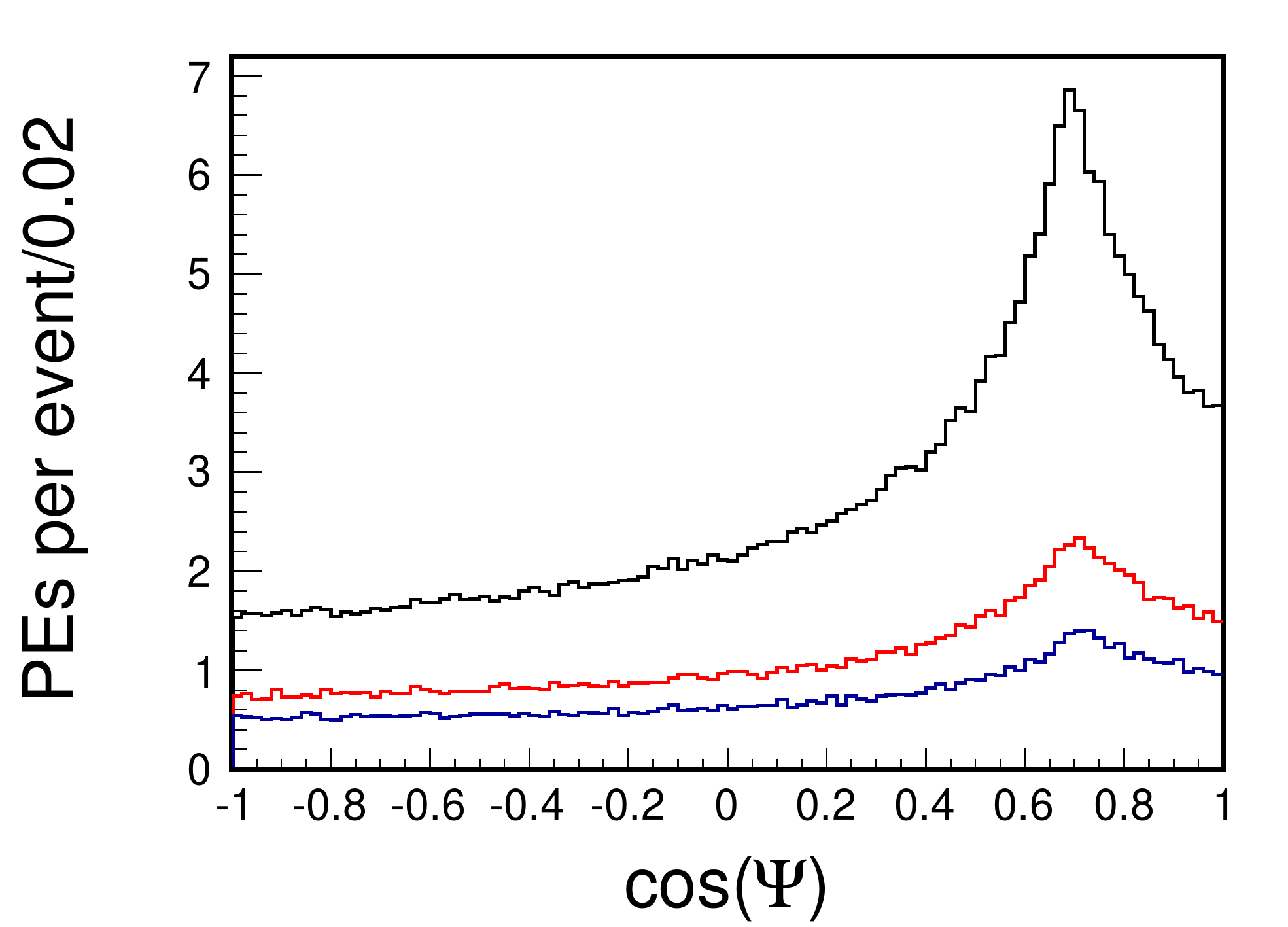}
        \caption[]{The angular distribution of photoelectron hits
        relative to the original electron direction, $\cos(\Psi) =
        x_{hit}/|\vec{r}_{hit}|$. Three energies are shown: 5~MeV (Black - Top), 2.1~MeV (Red - Middle), 1.4~MeV (Blue - Bottom). Each sample consists of 1000 events produced at the detector center. Default
        simulation settings are used and both Cherenkov and
        scintillation light are included. The $t\leq$ 34.0~ns cut is applied.} 
        \label{Cherenkov_cone}
        \end{center}
\end{figure}

Figure \ref{Cherenkov_cone} displays the angular distribution of PE
hits after the time cut. Although this time cut is a simplification of actual time
reconstruction effects, we can use it to indicate the spatial
distribution of hits in the early time window. The Cherenkov ring structure
can be clearly seen in the peak near 46\textdegree, demonstrating
that the directional signal conveyed by the Cherenkov photons is not
erased by scattering of the initial electrons even at 1.4~MeV.

When the 17-inch KamLAND PMTs\cite{tajimaMaster,kume_1983} (TTS =
1.28~ns) are used in the simulation, the broadening of the time
distributions leads to a strongly decreased ratio of Cherenkov over
scintillation light ($R_{C/S}=0.25$) for $t<34.0$~ns (see
figure \ref{time_plots_comparison} (b)). This shows that a
photodetector with a low TTS is critical for directionality reconstruction and
motivates the use of novel photodetector types.

\section{Detector wavelength response}
\label{detector_wavelength_response_sec} 
Since Cherenkov photons that pass through meters of
scintillator have on average longer wavelengths than scintillation
photons, a photodetector that is more sensitive at long wavelengths
increases not only the absolute number of PEs but also the ratio
between Cherenkov- and scintillation-induced PEs. We have run the simulation with the QE of an extended red-sensitive
GaAsP photocathode (Hamamatsu R3809U-63)\cite{Hamamatsu_R3899U}, but with the default TTS of 0.1~ns.
Figure \ref{time_plots_comparison}(c) shows the results for the
modified simulation with high QE in the red spectral region. The
higher absolute number of photoelectrons coming from Cherenkov light
(factor of $\approx$ 2) and the increased Cherenkov/scintillation
ratio ($R_{C/S}=1.01$) in the early time window would significantly improve the
directionality reconstruction.

\section{Scintillator emission spectrum}
\label{scintillator_emission_sec}
An alternative route towards increasing the separation in time between
Cherenkov and scintillation photon hits is the tuning of the
scintillator emission spectrum. Recently, the use of quantum dots
(QDs) in liquid scintillators has been studied as a possibility to
improve future large scale neutrino experiments\cite{qdot,qdot2}. One
major motivation for quantum-dot-doped scintillator is control of the
emission spectrum by tuning the size or composition of the quantum
dots. 

Quantum dots can also provide a mechanism for introducing an 
isotope for studying double-beta decay. The emission spectrum of commercial alloyed core/shell
CdS$_x$Se$_{1-x}$/ZnS quantum dots was measured in
Ref.\cite{qdot2}. This spectrum shows a symmetric peak centered
around 461~nm with FWHM = 29~nm.  In order to isolate the effect of
the different emission spectrum, the other simulation settings,
including the KamLAND absorption spectrum, were kept unchanged; we
find $R_{C/S}=0.17$ for the default 34.0~ns timing cut.  Compared to
the default case shown in figure \ref{time_plots_comparison}(a), the
separation is worse (as expected) because the scintillation light
wavelengths are longer than in the KamLAND emission spectrum.

However, advances in the production of commercial quantum dot samples
could yield quantum dots that have similar, single peak emission
shapes at shorter wavelengths. This case has been simulated using the
same spectral shape of the measured core-shell quantum dot emission
but shifted to shorter wavelengths such that the emission peak is
centered at 384~nm. This peak emission value has been measured for
other types of QDs, however with a much more pronounced 
tail\cite{qdot2}. The resulting PE time distribution shows improved
separation of Cherenkov and scintillation light compared to the
default simulation. After the 34.0~ns cut on the TTS-smeared PE time
we obtain a Cherenkov/scintillation ratio of $R_{C/S}=0.86$ (107 PE
from Cherenkov light and 124 PE from scintillation). The number of
Cherenkov-induced PEs after the time cut is unchanged while the number
of PEs coming from scintillation light is decreased due to the higher
average photon travel times.

\begin{figure}[tbh]
        \begin{center}
        \subfigure{\includegraphics[scale=0.375]{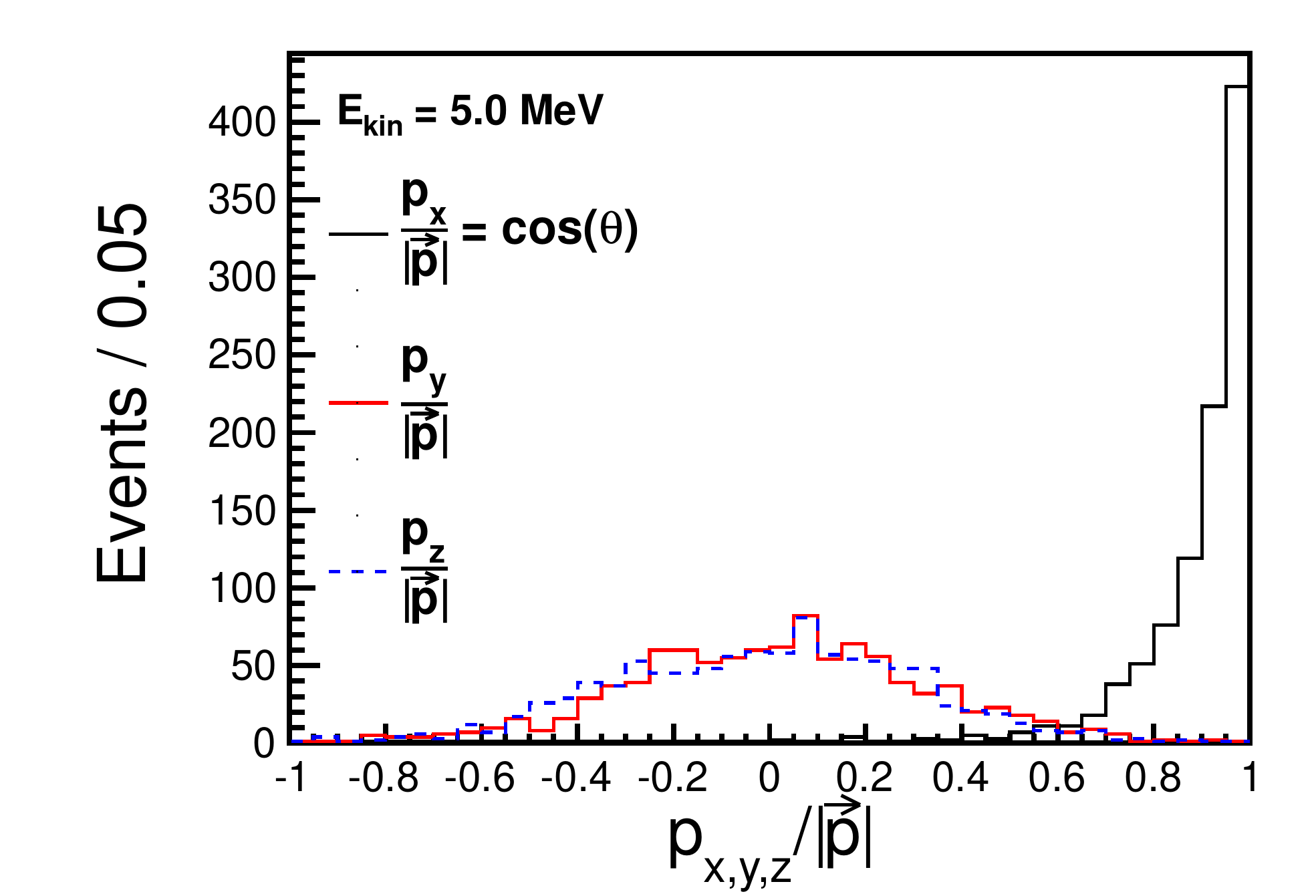}}
        \subfigure{\includegraphics[scale=0.375]{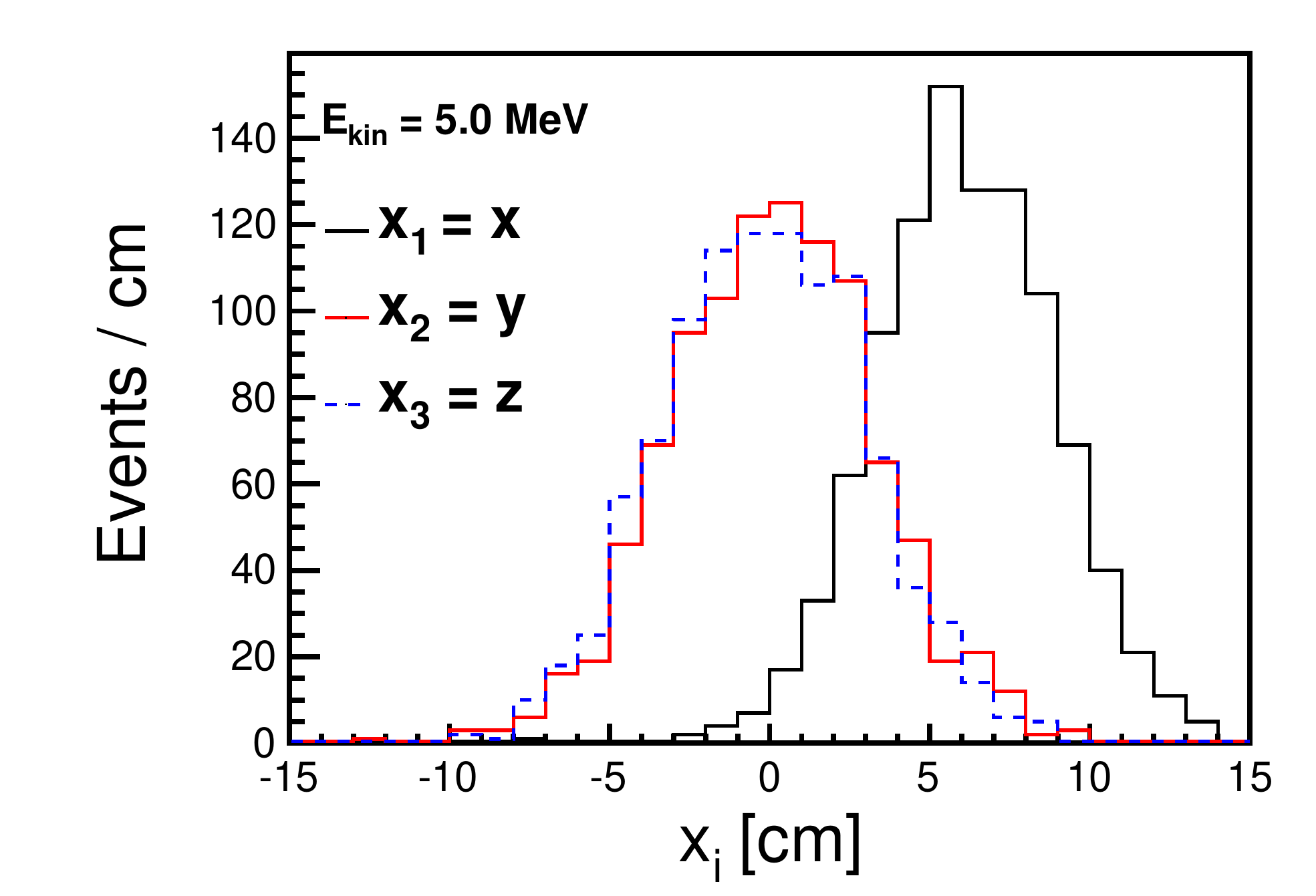}}
        \subfigure{\includegraphics[scale=0.375]{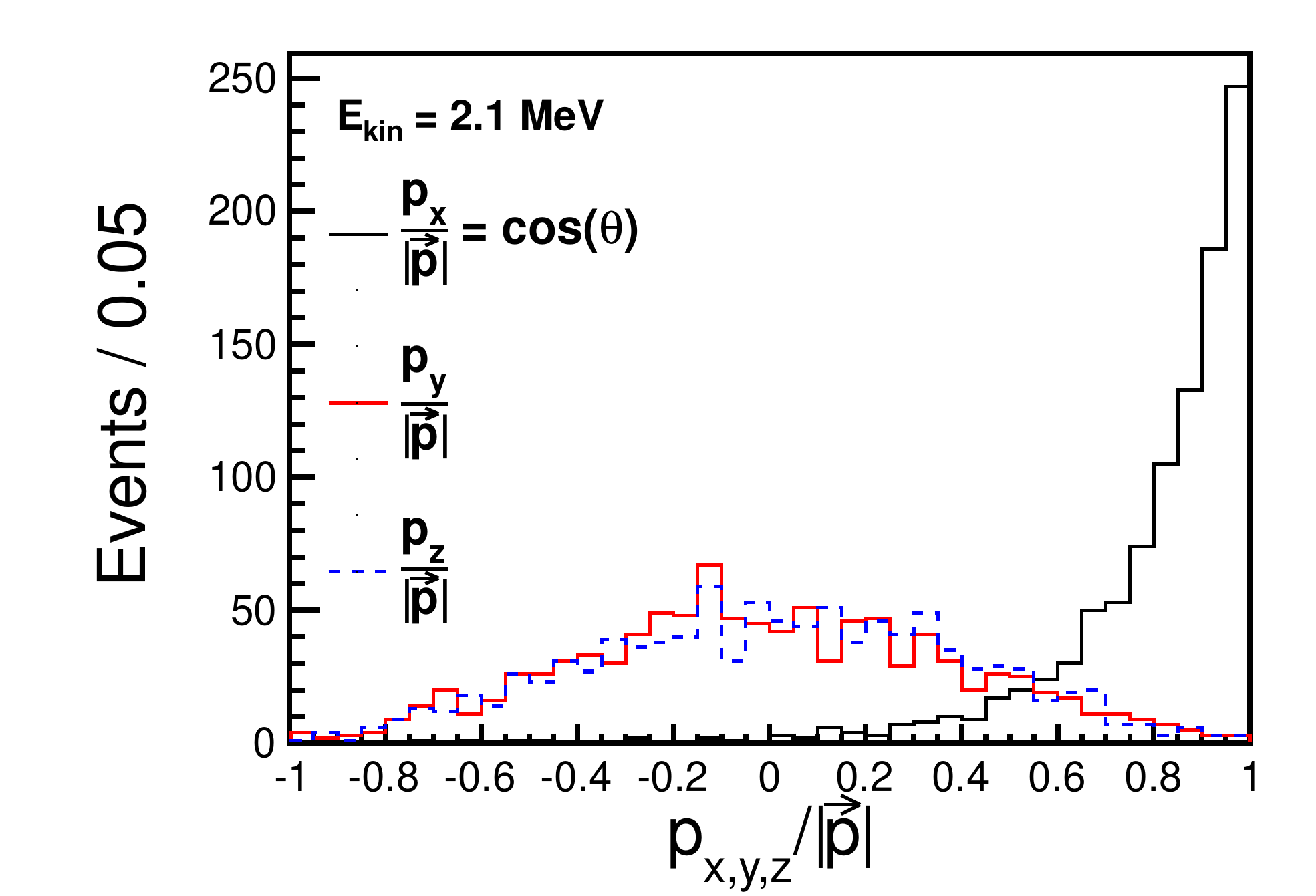}}
        \subfigure{\includegraphics[scale=0.375]{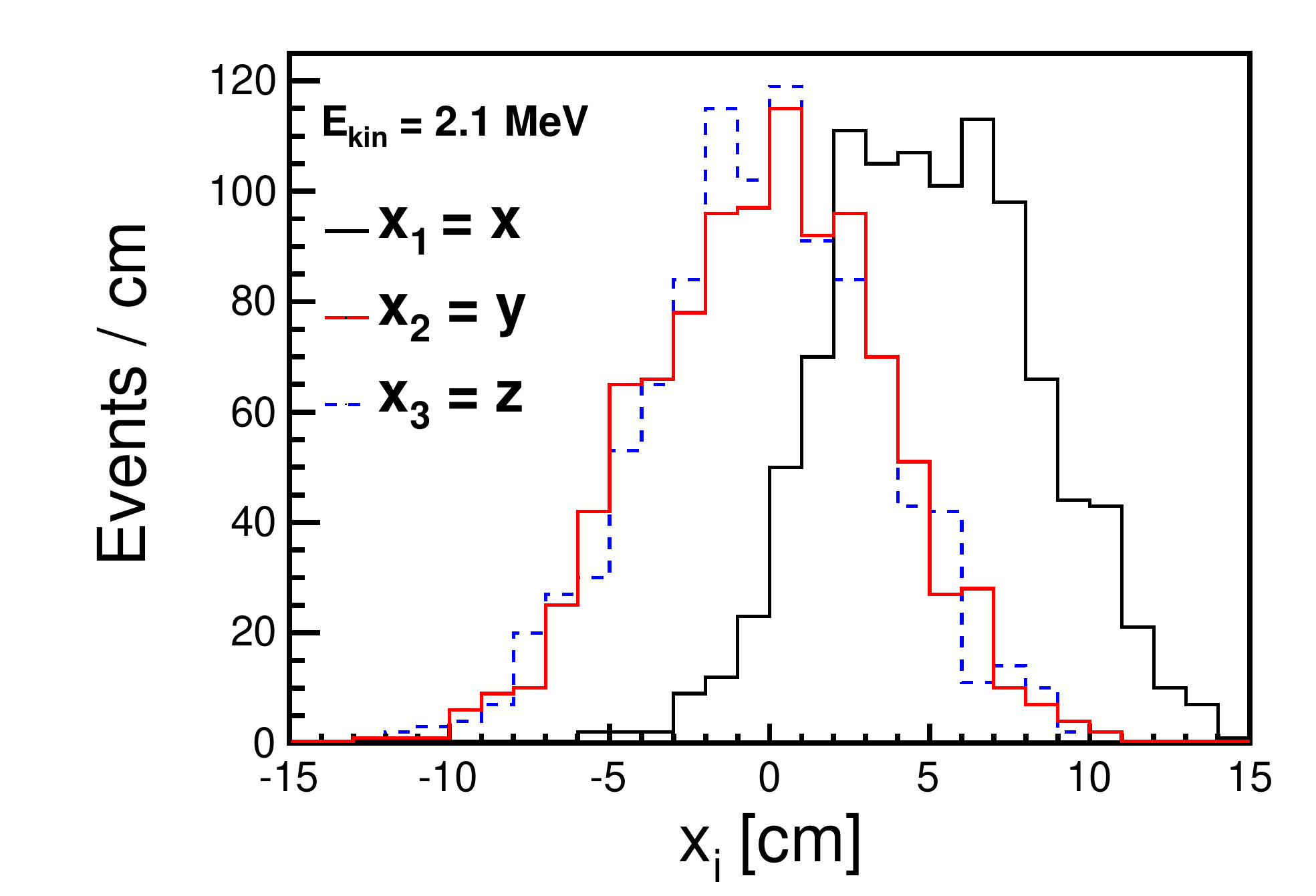}}
        \subfigure{\includegraphics[scale=0.375]{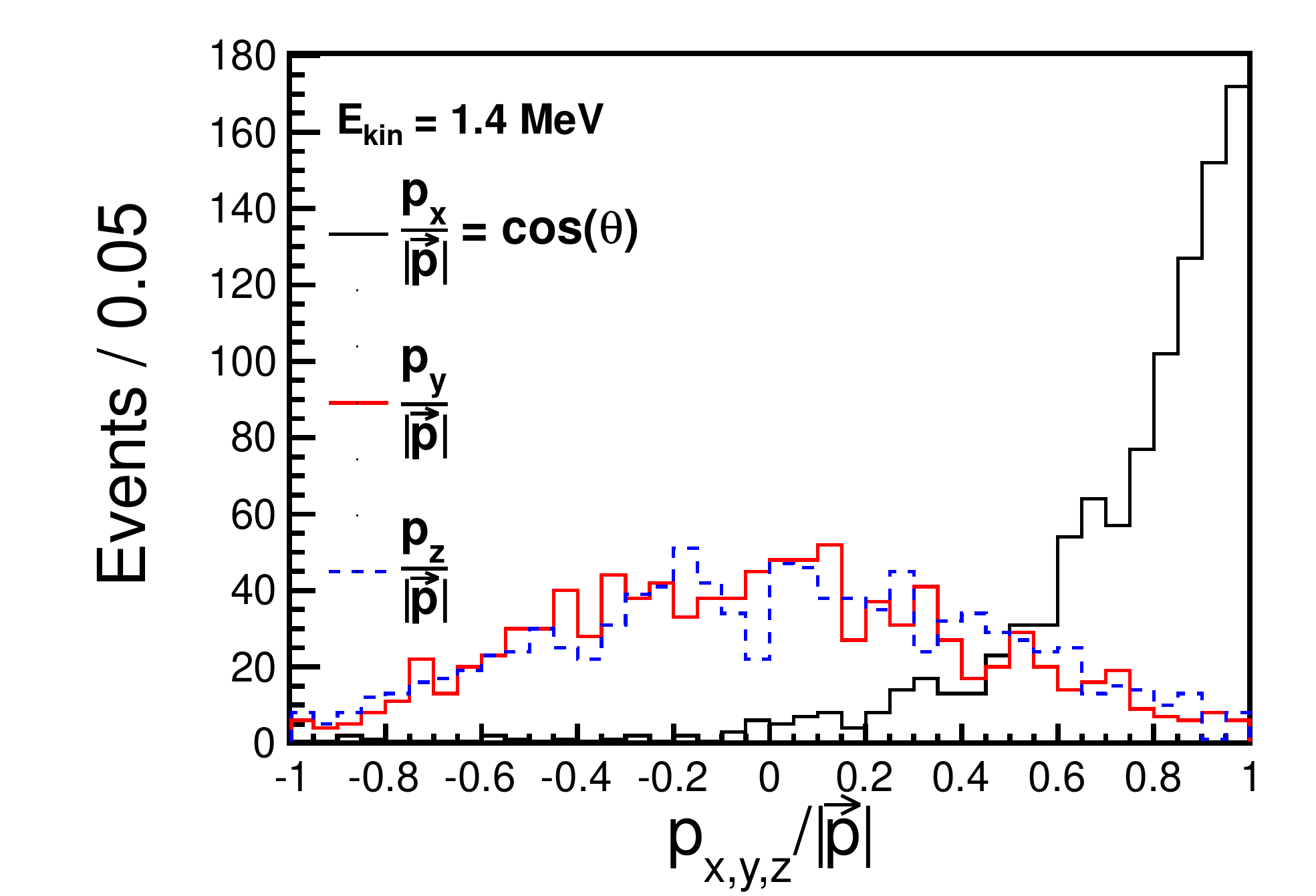}}
        \subfigure{\includegraphics[scale=0.375]{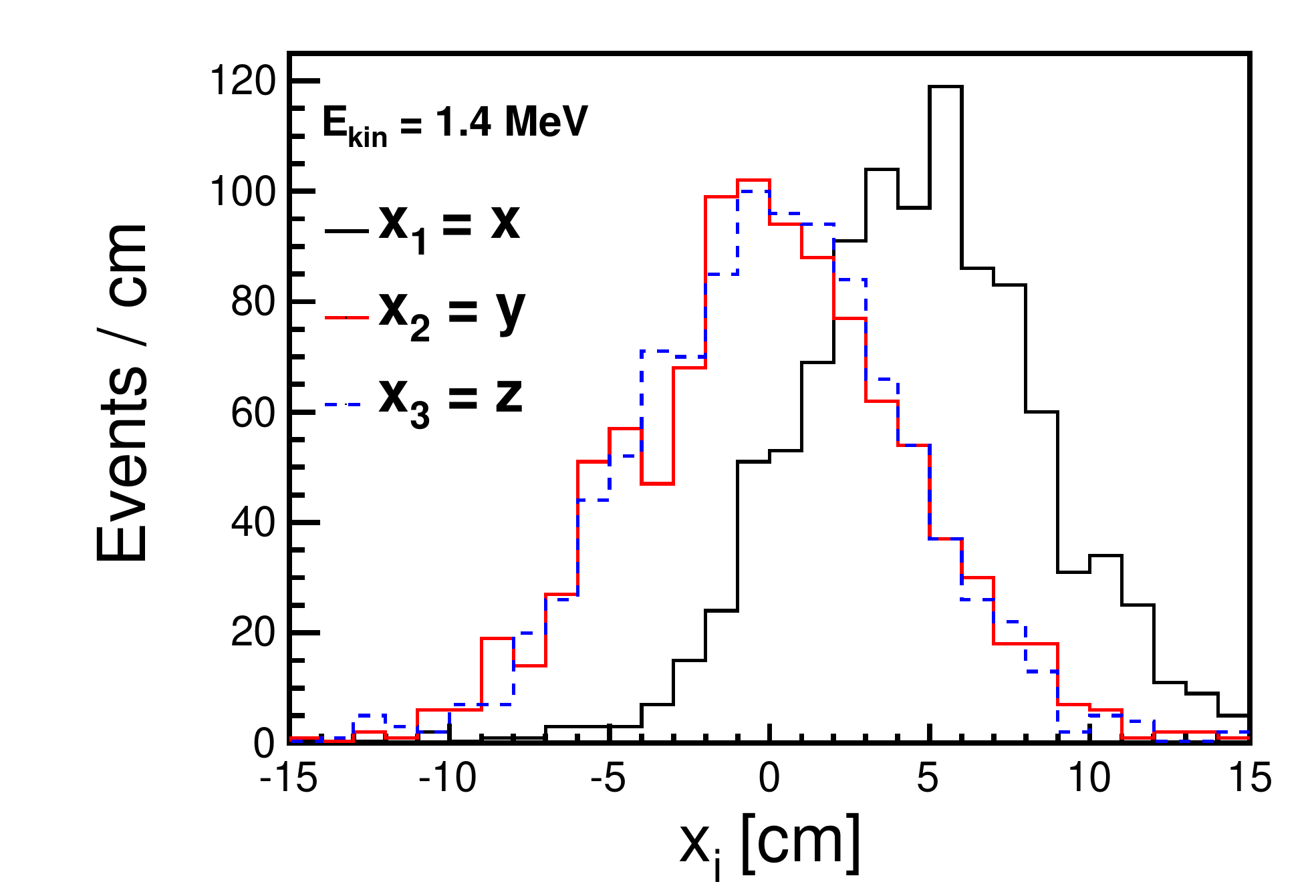}}
        \caption[]{\label{fig:reco} (Left) The reconstructed direction,
        $(p_x/|\vec{p}|, p_y/|\vec{p}|,
        p_z/|\vec{p}|)$, for the simulation of 1000 electrons. In the simulation the electrons are produced along the
        x-axis, $\vec{p}/|\vec{p}|$ = (1,0,0), and originate
        at the center of the 6.5m-radius detector, $\vec{r}$ =
        (0,0,0). Only photons with arrival time of $t<$ 34.0~ns are used
        in the reconstruction. The quantum efficiency of the bialkali
        photocathode is taken into account. (Right) The reconstructed
        vertex position, $(x,y,z)$, for the same simulation. From Top to Bottom, 5~MeV, 2.1~MeV and 1.4 MeV are shown.}
\end{center}
\end{figure}

\section{Reconstruction}
\label{reconstruction_sec}

The timing studies show that in the early time window, $t\leq$
34.0~ns, the ratio $R_{C/S}$ is high, improving the photoelectron hit selection. In this section, we apply
reconstruction tools for a water Cherenkov detector, WCSimAnalysis,
to the problem of reconstructing the position and direction of 5~MeV
electrons from this early light. WCSimAnalysis is a water Cherenkov
reconstruction package developed by the Long Baseline Neutrino
Experiment (LBNE) Collaboration\cite{Blake}. It provides a framework
for generic event cleaning, track reconstruction, and particle
identification, and comes equipped with variety of pre-built
algorithms. It is continuing to be expanded using new track-fitting
techniques for water Cherenkov detectors\cite{Sanchez2012525} based on
advanced photosensors with sub-cm imaging capabilities and timing
resolutions below 100 picoseconds.

To start, we are neglecting the effects of position dependence. In future work the arrival times can be corrected by the time of flight from the reconstructed vertex and the position and direction fitted simultaneously. For isotropic light, the vertex reconstruction uncertainty leads to an additional smearing of the arrival time distribution of 0.15~ns for every 3~cm of position reconstruction uncertainty. For directional light, the vertex reconstruction uncertainty leads to an effective shift of the arrival time distribution. This would change the ideal time cut by 0.15~ns  for every 3~cm of position reconstruction uncertainty. We have studied other time cuts from 33~ns to 34.5~ns and the reconstruction remains reasonable. 

The results presented in this paper rely on a simple vertex
reconstruction algorithm, commonly known as a ``point
fit''\cite{SuperKalgo}. It assumes that all of the scintillation and
Cherenkov light is emitted from a single point in space-time
$(x_0,y_0,z_0,t_0)$. In actuality, the light is emitted along a multi-scattered electron track. However, at the energies
discussed in this paper, the extent of this track is small (a few cm)
compared to the scale of the detector (R=6.5 meters) that sets the typical
photon transit distances.

The first step of the reconstruction process relies on exact numerical
calculations of vertex candidates from quadruplets of hits. Given a
single point source, we need four constraints to solve for the four
unknowns of the vertex $(x,y,z,t_0)$\cite{Smy}. This approach
would provide an exact solution in the case of four prompt,
un-scattered photons originating from a common point. However, many of
these randomly chosen quadruplets will produce anomalous solutions due
to `real world' effects such as delayed emission and deviations from the
point-like geometry. Nonetheless, we found that any chosen subset of
400 quadruplets was a sufficiently large ensemble to assure that some
solutions will be close to the true vertex.

Once a set of vertex candidates has been found, we test the goodness
of each vertex and select the one that best fits the full ensemble of
photon hits. The goodness of fit is determined based on the
distribution of an observable known as the ``point time
residual''\cite{SuperKalgo}. The point time residual is calculated by first choosing a hypothesis for the vertex position and $t_0$ for the event. The goodness of fit for these values  is then calculated by taking the difference between the measured and predicted times of each photon hit, using a single effective speed of light in the scintillator. The width of the time residual
distribution over all hits is minimized when the hypothesized vertex
is near the true vertex. Based on this figure of merit, we select the
vertex with the narrowest time residual distribution from among the
400 candidates.

The direction of the electron track is then determined by taking the
centroid of all vectors pointing from the fitted vertex to the hits on
the detector. Since the Cherenkov light is highly directional, and
since the timing cut enhances the purity of the Cherenkov light in the
sample, this calculation provides a good measure of the track
direction. 

For the purpose of testing the reconstruction algorithm we use 1000
simulated electrons with an energy of 5~MeV; lower energies are studied in the next section.
The electrons are simulated at the center of the detector, $\vec{r}$ = (0,0,0), along
the x-axis, $\vec{p}/|\vec{p}|$ = (1,0,0). Figure \ref{fig:reco} (Top)
shows the vertex reconstruction. The vertex is reasonably well-reconstructed around the center of the detector, $\vec{r}$ = (0,0,0),
except along the x-axis. The RMS values of the distributions for all three
reconstructed coordinates are smaller than 3.5~cm. The shift along
the x-axis is due to two effects for which the reconstruction has to
use average values rather than the unknown true value for each
hit: the wavelength and hence the speed of the light in the medium,
and the point of emission for each of the photons, reconstructed
as coming from a common point. The reconstruction of the direction
also is shown in figure \ref{fig:reco} (Top). It shows that for the majority
of the events the initial electron direction is reconstructed well.
This is a promising result given the simplicity of the algorithms.

\section{Energy dependence}
\label{edep_section}
In the previous sections we presented results on single 5~MeV electrons such as might be observed in neutrino-electron scattering.
In this section we study two lower energies, 1.4~MeV and 2.1~MeV, appropriate for searches for $0\nu\beta\beta$. These
energies correspond to Q/2 for the double-beta decay of
$^{116}$Cd and $^{48}$Ca, respectively\cite{cd1,biller}. The isotope
$^{116}$Cd was chosen because of its potential use in quantum-dot-doped
scintillators\cite{qdot,qdot2} and $^{48}$Ca was chosen
to cover the Q-value range of $0\nu\beta\beta$ candidate isotopes.

\begin{figure}
        \begin{center}
        \includegraphics[scale=0.5]{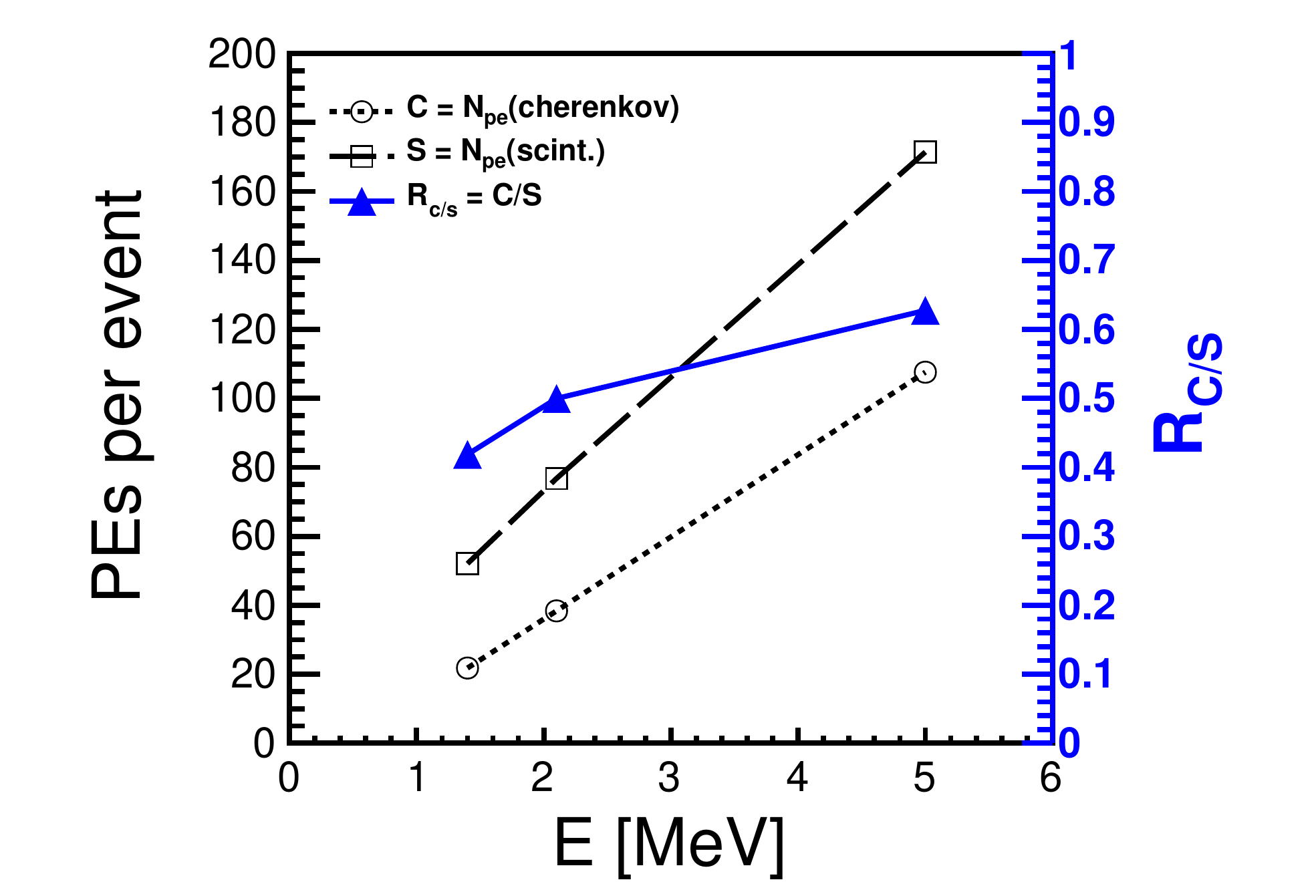}
        \caption[]{The energy dependence of the mean number of PEs after the 34.0~ns time cut is shown for Cherenkov-induced PEs (black open circles, dotted line) and scintillation-induced PEs (black open squares, dashed line). The ratio between the mean number of Cherenkov-induced and scintillation-induced PEs is shown as blue filled triangles, values are given on the right y-axis. The statistical errors are too small to be seen. \label{Edep_NPE}}
        \end{center}
\end{figure}

The two additional simulation sets with 1.4 MeV and 2.1 MeV electrons were
generated using the default simulation configuration described in section
\ref{sim_section}. The PE time distribution for the default settings is shown
in figure \ref{time_plots_comparison} (a) for 5~MeV electrons. The shapes
of the scintillation and Cherenkov spectra are similar for the lower energies (not
shown here). In figure \ref{Edep_NPE}, the energy-dependent mean number of
PEs per event after the 34.0~ns time cut is shown for Cherenkov-induced and
scintillation-induced PEs, as well as their ratio $R_{C/S}$. The mean number
of PEs from Cherenkov (scintillation) light is 21.8 (52.1), 38.4 (76.8)
and 108 (171) for electron energies of 1.4~MeV, 2.1~MeV and 5~MeV, respectively. This gives the
ratios $R_{C/S}$ = 0.42, 0.50 and 0.63: The decrease in Cherenkov-induced
PEs is stronger than the decrease in scintillation-induced PEs as the energy
is lowered.

\begin{figure}
        \begin{center}
        \includegraphics[scale=0.4]{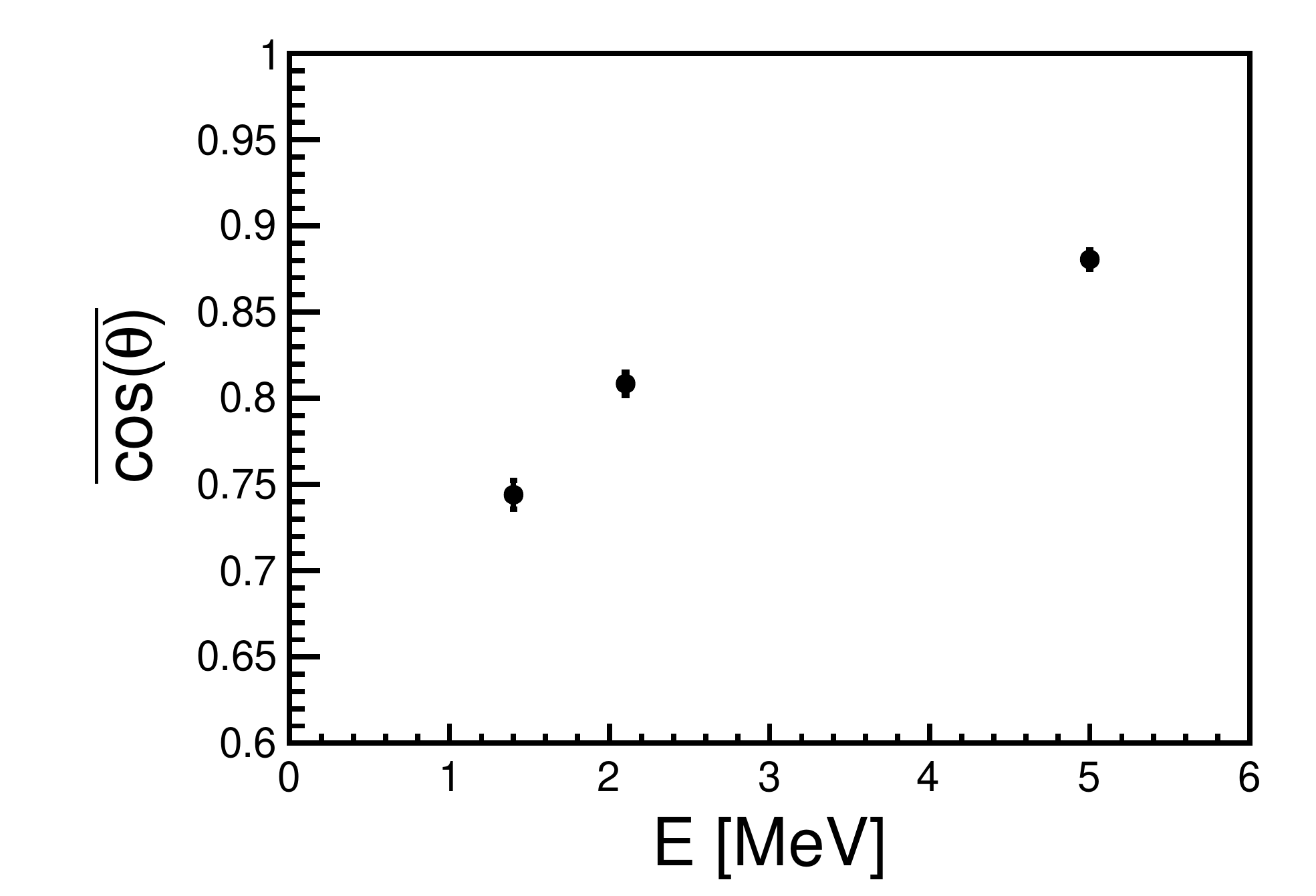}
        \caption[]{Mean cosine of the angle between the true initial electron direction and the reconstructed direction, as a function of the electron energy. For each energy 1000 events have been simulated. Statistical errors are
shown. \label{Edep_angle}}
        \end{center}
\end{figure}

The reconstruction algorithms outlined in section \ref{reconstruction_sec} have also been applied
to the simulations at lower energies. Figure \ref{fig:reco} (Middle) shows the results for 2.1~MeV and figure \ref{fig:reco} (Bottom) shows the results for 1.4~MeV. Most events are still reconstructed well, despite the lower
number of PEs and the decreased $R_{C/S}$. For 1.4~MeV electrons, the RMS values of the distributions for all three
reconstructed coordinates are smaller than 4.5~cm. The mean cosine of the angle between the true direction
and the reconstructed direction for different energies of the initial electron is shown in figure
\ref{Edep_angle}. The direction reconstruction performance is still promising for energies as low as 1.4~MeV.

\section{Conclusions}

We have developed a technique to separate scintillation and Cherenkov light to reconstruct direction of electrons with energies 5~MeV, 2.1~MeV and 1.4~MeV in a liquid scintillator detector. These energies have been chosen to represent typical neutrino-electron elastic scattering energies and
$0\nu\beta\beta$ energies. The Cherenkov threshold for an electron in a typical liquid scintillator is $\sim$0.2~MeV.

While scintillation light is isotropic, Cherenkov light with wavelengths above the absorption cutoff $\sim$370~nm carries the information about the direction of the electrons. All light shorter than this gets absorbed and re-emitted isotropically as a part of the scintillation process. On average scintillation light is delayed with respect to the direct Cherenkov light due to chromatic dispersion and the finite time of the scintillation processes; the early light thus contains directional information.

Using a Geant4~simulation of a spherical detector with radius of 6.5~m and photodetectors with TTS of 0.1~ns, we have shown that for electrons originated at the center a time cut on the early light is effective at isolating the directional light, improving the ratio of Cherenkov light to scintillation light from 0.02 to 0.63. This ratio is degraded by a factor of 2.5 if current photodetectors with $\sim$1~ns resolution were used. This can be improved by factors of 1.6 or 1.4 if more red-sensitive photodetectors or scintillators with narrower emission spectra are used.

Reconstruction algorithms developed for water Cherenkov detectors have been applied to this early light. The algorithms are able to converge on reasonable reconstructed vertices and directions for all simulated energies: 5~MeV, 2.1~MeV, 1.4~MeV. As expected, the reduction in photon statistics with lower energies leads to a broadening of the reconstructed vertex by $\sim$1~cm. Similarly, the mean $\cos(\theta)$ between the reconstructed and initial directions drops from 0.88 to 0.74.

This technique is promising, and we plan to continue work on the topic. The ability to reconstruct direction in kiloton-scale scintillation detectors would expand capabilities of neutrino experiments, especially those also searching for neutrino-less double-beta decay. More generally, this technique could be applied wherever scintillation-based detectors are used.



\acknowledgments
The authors thank Andrew Blake at University of Cambridge for his work
authoring the WCSimAnalysis code. The authors thank the neutrino
reconstruction group at Iowa State, particularly Mayly Sanchez, Ioana
Anghel, and Tian Xin, for their continued work in developing the
WCSimAnalysis algorithms and for their insights and expertise
regarding issues related to Cherenkov reconstruction with
fast-timing. The authors also thank Michael Smy for his development of
the quadruplet-based vertex-finding method. L. Winslow would like to
thank Janet Conrad for many useful discussions on the topic, and
Katsushi Arisaka for discussions on the possible reach of traditional PMTs
and the characteristics of HPDs. C. Aberle and L. Winslow are
supported by funds from University of California Los Angeles. The work
at the University of Chicago is partially supported by DOE
contract DE-SC0008172 and NSF grant PHY-1066014. Matthew Wetstein gratefully
acknowledges support by the Grainger Foundation.

\newpage

\bibliographystyle{JHEP}
\bibliography{DirectionBibliography_v1} 

\providecommand{\href}[2]{#2}\begingroup\raggedright\begin{thebibliography}{10}

\bibitem{kam2013}
{\bf KamLAND} Collaboration, A.~Gando et~al., {\it {Reactor On-Off Antineutrino
  Measurement with KamLAND}},  \href{http://xxx.lanl.gov/abs/1303.4667}{{\tt
  arXiv:1303.4667}}.

\bibitem{borexino}
{\bf Borexino} Collaboration, G.~Bellini et~al., {\it {Precision measurement of
  the 7Be solar neutrino interaction rate in Borexino}},  {\em Phys. Rev.
  Lett.} {\bf 107} (2011) 141302,
  [\href{http://xxx.lanl.gov/abs/1104.1816}{{\tt arXiv:1104.1816}}].

\bibitem{dbtwo}
{\bf Daya Bay} Collaboration, F.~An et~al., {\it {Improved Measurement of
  Electron Antineutrino Disappearance at Daya Bay}},  {\em Chin. Phys.} {\bf
  C37} (2013) 011001, [\href{http://xxx.lanl.gov/abs/1210.6327}{{\tt
  arXiv:1210.6327}}].

\bibitem{dctwo}
{\bf Double Chooz} Collaboration, Y.~Abe et~al., {\it {Reactor electron
  antineutrino disappearance in the Double Chooz experiment}},  {\em Phys.
  Rev.} {\bf D86} (2012) 052008, [\href{http://xxx.lanl.gov/abs/1207.6632}{{\tt
  arXiv:1207.6632}}].

\bibitem{dchydrogen}
{\bf Double Chooz} Collaboration, Y.~Abe et~al., {\it {First Measurement of
  $\theta_{13}$ from Delayed Neutron Capture on Hydrogen in the Double Chooz
  Experiment}},  {\em Phys. Lett.} {\bf B723} (2013) 66,
  [\href{http://xxx.lanl.gov/abs/1301.2948}{{\tt arXiv:1301.2948}}].

\bibitem{reno}
{\bf RENO} Collaboration, J.~K. Ahn et~al., {\it {Observation of Reactor
  Electron Antineutrinos Disappearance in the RENO Experiment}},  {\em Phys.
  Rev. Lett.} {\bf 108} (2012) 191802.

\bibitem{juno}
Y.-F. Li, J.~Cao, Y.~Wang, and L.~Zhan, {\it {Unambiguous Determination of the
  Neutrino Mass Hierarchy Using Reactor Neutrinos}},
  \href{http://xxx.lanl.gov/abs/1303.6733}{{\tt arXiv:1303.6733}}.

\bibitem{reno50}
{\it {RENO-50 - International Workshop on toward Neutrino Mass Hierarchy}},
  June, 2009.

\bibitem{isodarscatt}
J.~Conrad, M.~Shaevitz, I.~Shimizu, J.~Spitz, M.~Toups, and L.~Winslow, {\it
  {\it Precision $\bar{\nu}_{e}$-electron Scattering Measurements with IsoDAR
  to Search for New Physics}},  \href{http://xxx.lanl.gov/abs/1307.5081}{{\tt
  arXiv:1307.5081}}. In preparation, for submission to Phys. Rev. D.

\bibitem{isodar}
{\bf IsoDAR} Collaboration, A.~Bungau et~al., {\it {Proposal for an Electron
  Antineutrino Disappearance Search Using High-Rate $^{8}$Li Production and
  Decay}},  {\em Phys. Rev. Lett.} {\bf 109} (2012) 141802,
  [\href{http://xxx.lanl.gov/abs/1205.4419}{{\tt arXiv:1205.4419}}].

\bibitem{nist}
K.~Heeger, B.~Littlejohn, and H.~Mumm, {\it {Multiple Detectors for a
  Short-Baseline Neutrino Oscillation Search Near Reactors}},
  \href{http://xxx.lanl.gov/abs/1307.2859}{{\tt arXiv:1307.2859}}.

\bibitem{nucifer}
A.~Porta et~al., {\it {Reactor Neutrino Detection for Non-Proliferation With
  the NUCIFER Experiment}},  {\em Nuclear Science, IEEE Transactions on} {\bf
  57} (2010) 2732--2739.

\bibitem{songs}
N.~Bowden et~al., {\it {Experimental results from an antineutrino detector for
  cooperative monitoring of nuclear reactors}},  {\em Nucl. Instrum. Meth.}
  {\bf A572} (2007) 985 -- 998.

\bibitem{leptogenesis}
M.~Fukugita and T.~Yanagida, {\it Barygenesis without grand unification},  {\em
  Physics Letters B} {\bf 174} (1986), no.~1 45 -- 47.

\bibitem{KZ0nu}
{\bf KamLAND-Zen} Collaboration, A.~Gando et~al., {\it {Limit on Neutrinoless
  $\beta\beta$ Decay of Xe-136 from the First Phase of KamLAND-Zen and
  Comparison with the Positive Claim in Ge-76}},  {\em Phys. Rev. Lett.} {\bf
  110} (2013) 062502, [\href{http://xxx.lanl.gov/abs/1211.3863}{{\tt
  arXiv:1211.3863}}].

\bibitem{gerda2013}
{\bf GERDA} Collaboration, M.~Agostini et~al., {\it {Results on neutrinoless
  double beta decay of 76Ge from GERDA Phase I}},  {\em Phys.Rev.Lett.} {\bf
  111} (2013) 122503, [\href{http://xxx.lanl.gov/abs/1307.4720}{{\tt
  arXiv:1307.4720}}].

\bibitem{Alessandria:2011rc}
{\bf CUORE} Collaboration, F.~Alessandria et~al., {\it {Sensitivity of CUORE to
  Neutrinoless Double-Beta Decay}},
  \href{http://xxx.lanl.gov/abs/1109.0494}{{\tt arXiv:1109.0494}}.

\bibitem{SuperNEMO}
{\bf SuperNEMO} Collaboration, R.~Arnold et~al., {\it {Probing New Physics
  Models of Neutrinoless Double Beta Decay with SuperNEMO}},  {\em Eur.Phys.J.}
  {\bf C70} (2010) 927--943, [\href{http://xxx.lanl.gov/abs/1005.1241}{{\tt
  arXiv:1005.1241}}].

\bibitem{EXO2012}
{\bf EXO} Collaboration, M.~Auger et~al., {\it {Search for Neutrinoless
  Double-Beta Decay in $^{136}\mathrm{Xe}$ with EXO-200}},  {\em Phys. Rev.
  Lett.} {\bf 109} (Jul, 2012) 032505.

\bibitem{NEXTsipm}
{\bf NEXT} Collaboration, V.~Alvarez et~al., {\it {Operation and first results
  of the NEXT-DEMO prototype using a silicon photomultiplier tracking array}},
  {\em JINST} {\bf 8} (2013) P09011,
  [\href{http://xxx.lanl.gov/abs/1306.0471}{{\tt arXiv:1306.0471}}].

\bibitem{john}
J.~G. Learned, {\it {High Energy Neutrino Physics with Liquid Scintillation
  Detectors}},  \href{http://xxx.lanl.gov/abs/0902.4009}{{\tt
  arXiv:0902.4009}}.

\bibitem{Gotthard}
R.~Luscher et~al., {\it {Search for beta beta decay in Xe-136: New results from
  the Gotthard experiment}},  {\em Phys.Lett.} {\bf B434} (1998) 407--414.

\bibitem{newphysics0nuBB}
A.~Ali, A.~Borisov, and D.~Zhuridov, {\it {Probing new physics in the
  neutrinoless double beta decay using electron angular correlation}},  {\em
  Phys.Rev.} {\bf D76} (2007) 093009,
  [\href{http://xxx.lanl.gov/abs/0706.4165}{{\tt arXiv:0706.4165}}].

\bibitem{chooz}
{\bf CHOOZ} Collaboration, M.~Apollonio et~al., {\it {Search for neutrino
  oscillations on a long baseline at the CHOOZ nuclear power station}},  {\em
  Eur.Phys.J.} {\bf C27} (2003) 331--374,
  [\href{http://xxx.lanl.gov/abs/hep-ex/0301017}{{\tt hep-ex/0301017}}].

\bibitem{dcDirection}
K.~A. Hochmuth, M.~Lindner, and G.~G. Raffelt, {\it {Exploiting the directional
  sensitivity of the Double Chooz near detector}},  {\em Phys.Rev.} {\bf D76}
  (2007) 073001, [\href{http://xxx.lanl.gov/abs/0704.3000}{{\tt
  arXiv:0704.3000}}].

\bibitem{birks_book}
{J.B. Birks}, {\em {The Theory and Practice of Scintillation Counting}}.
\newblock {Pergamon Press}, {1964}.

\bibitem{Cherenkov34}
P.~A. Cherenkov, {\it {Visible emission of clean liquids by action of gamma
  radiation}},  {\em Doklady Akademii Nauk SSSR} {\bf 2} (1934) 451.

\bibitem{group_velocity_article}
A.~M. Steinberg, P.~G. Kwiat, and R.~Y. Chiao, {\it Dispersion cancellation in
  a measurement of the single-photon propagation velocity in glass},  {\em
  Phys. Rev. Lett.} {\bf 68} (1992) 2421--2424.

\bibitem{pdg_review_2012}
{\bf Particle Data Group} Collaboration, J.~Beringer et~al., {\it {Review of
  Particle Physics (RPP)}},  {\em Phys. Rev.} {\bf D86} (2012) 010001.

\bibitem{tamm1939}
I.~Tamm, {\it {Radiation Emitted by Uniformly Moving Electrons}},  {\em J.
  Phys. U.S.S.R.} {\bf 1} (1939) 439.

\bibitem{bandv}
F.~Boehm and P.~Vogel, {\em Physics of Massive Neutrinos}.
\newblock Cambridge University Press, 1992.

\bibitem{tabledbb}
V.~Tretyak and Y.~Zdesenko, {\it Tables of double beta decay data},  {\em
  Atomic Data and Nuclear Data Tables} {\bf 61} (1995) 43 -- 90.

\bibitem{minfang}
M.~Yeh, Y.~Williamson, and R.~L. Hahn, {\it {Metal-loaded liquid scintillators
  for neutrino experiments}},  {\em J. Phys. Conf. Ser.} {\bf 136} (2008)
  042054.

\bibitem{nd1}
I.~Barabanov et~al., {\it {A Nd-loaded liquid organic scintillator for the
  experiment aimed at measuring double beta decay}},  {\em Instrum. Exp. Tech.}
  {\bf 55} (2012) 545--550.

\bibitem{zr1}
Y.~Fukuda, S.~Moriyama, and I.~Ogawa, ``{Development of liquid scintillator
  containing a zirconium complex for neutrinoless double beta decay
  experiment}.'' article in press, Nucl. Instrum. Meth. A, 2013.

\bibitem{mo1}
V.~Gehman, P.~Doe, R.~Robertson, D.~Will, H.~Ejiri, and R.~Hazama, {\it
  {Solubility, Light Output and Energy Resolution Studies of Molybdenum-Loaded
  Liquid Scintillators}},  {\em Nucl. Instrum. Meth.} {\bf A622} (2010)
  602--607, [\href{http://xxx.lanl.gov/abs/0911.2198}{{\tt arXiv:0911.2198}}].

\bibitem{qdot}
L.~Winslow and R.~Simpson, {\it Characterizing quantum-dot-doped liquid
  scintillator for applications to neutrino detectors},  {\em Journal of
  Instrumentation} {\bf 7} (2012) P07010.

\bibitem{cd1}
G.~Bellini et~al., {\it {High sensitivity 2 beta decay study of Cd-116 and
  Mo-100 with the BOREXINO counting test facility (CAMEO project)}},  {\em Eur.
  Phys. J.} {\bf C19} (2001) 43--55,
  [\href{http://xxx.lanl.gov/abs/nucl-ex/0007012}{{\tt nucl-ex/0007012}}].

\bibitem{biller}
S.~D. Biller, {\it {Probing Majorana neutrinos in the regime of the normal mass
  hierarchy}},  {\em Phys. Rev.} {\bf D87} (2013) 071301,
  [\href{http://xxx.lanl.gov/abs/1306.5654}{{\tt arXiv:1306.5654}}].

\bibitem{sn1}
{\bf KIMS} Collaboration, M.~Hwang et~al., {\it {Development of tin-loaded
  liquid scintillator for the double beta decay experiment}},  {\em Nucl.
  Instrum. Meth.} {\bf A570} (2007) 454--458.

\bibitem{qdot2}
C.~Aberle, J.~Li, S.~Weiss, and L.~Winslow, {\it {Optical properties of
  quantum-dot-doped liquid scintillators}},  {\em Journal of Instrumentation}
  {\bf 8} (2013) P10015, [\href{http://xxx.lanl.gov/abs/1307.4742}{{\tt
  arXiv:1307.4742}}].

\bibitem{geant4one}
{\bf GEANT4} Collaboration, S.~Agostinelli et~al., {\it {GEANT4: A Simulation
  toolkit}},  {\em Nucl. Instrum. Meth.} {\bf A506} (2003) 250--303.

\bibitem{geant4two}
J.~Allison et~al., {\it Geant4 developments and applications},  {\em Nuclear
  Science, IEEE Transactions on} {\bf 53} (2006) 270--278.

\bibitem{kamland2003}
{\bf KamLAND} Collaboration, K.~Eguchi et~al., {\it {First results from
  KamLAND: Evidence for reactor anti-neutrino disappearance}},  {\em Phys. Rev.
  Lett.} {\bf 90} (2003) 021802,
  [\href{http://xxx.lanl.gov/abs/hep-ex/0212021}{{\tt hep-ex/0212021}}].

\bibitem{tajimaMaster}
O.~Tajima, {\it {\it Development of Liquid Scintillator for a Large Size
  Neutrino Detector}},  Master's thesis, Tohoku University, 2000.

\bibitem{OlegThesis}
O.~Perevozchikov, {\em {\it Search for electron antineutrinos from the sun with
  KamLAND detector}}.
\newblock PhD thesis, University of Tennessee, 2009.

\bibitem{tajimaThesis}
O.~Tajima, {\em {\it Measurement of Electron Anti-Neutrino Oscillation
  Parameters with a Large Volume Liquid Scintillator Detector, KamLAND}}.
\newblock PhD thesis, Tohoku University, 2003.

\bibitem{ChrisThesis}
C.~Grant, {\em {A Monte Carlo Approach to $^{7}$Be Solar Neutrino Analysis with
  KamLAND}}.
\newblock PhD thesis, University of Alabama, 2012.

\bibitem{Wurm:2010ad}
M.~Wurm, F.~von Feilitzsch, M.~Goeger-Neff, M.~Hofmann, T.~Lachenmaier, et~al.,
  {\it {Optical Scattering Lengths in Large Liquid-Scintillator Neutrino
  Detectors}},  {\em Rev.Sci.Instrum.} {\bf 81} (2010) 053301,
  [\href{http://xxx.lanl.gov/abs/1004.0811}{{\tt arXiv:1004.0811}}].

\bibitem{Hamamatsu_R7081}
{\it {Hamamatsu {P}hotonics {K}.{K}., {L}arge Photocathode Area Photomultiplier
  Tubes (data sheet, including {R}7081)}},  {accessed July}, 2013.
\newblock {http://www.hamamatsu.com/resources/pdf/etd/LARGE\_
  AREA\_PMT\_TPMH1286E05.pdf}.

\bibitem{geant4scatt}
{\it {Electromagnetic Standard Physics Working Group}},  {accessed December},
  {2013}.
\newblock
  {http://www.geant4.org/geant4/collaboration/working\_groups/electromagnetic/index.shtml}.

\bibitem{Adams:2013nva}
B.~Adams et~al., {\it {Measurements of the gain, time resolution, and spatial
  resolution of a 20x20cm$^2$ MCP-based picosecond photo-detector}},  {\em
  Nucl.Instrum.Meth.} {\bf A732} (2013) 392--396.

\bibitem{RSI_paper}
B.~Adams et~al., {\it Invited article: A test-facility for large-area
  microchannel plate detector assemblies using a pulsed sub-picosecond laser},
  {\em Review of Scientific Instruments} {\bf 84} (2013) 061301.

\bibitem{PSEC4_paper}
E.~Oberla, J.~Genat, H.~Grabas, H.~Frisch, K.~Nishimura, and G.~Varner, ``{ A
  15 GSa/s, 1.5 GHz Bandwidth Waveform Digitizing ASIC}.'' submitted to Nucl.
  Instrum. Meth. A, 2013.

\bibitem{anode_paper}
H.~Grabas, R.~Obaid, E.~Oberla, H.~Frisch, J.-F. Genat, R.~Northrop, F.~Tang,
  D.~McGinnis, B.~Adams, and M.~Wetstein, {\it {RF strip-line anodes for Psec
  large-area MCP-based photodetectors}},  {\em Nucl.Instrum.Meth.} {\bf A 711}
  (2013) 124 -- 131.

\bibitem{hpdThesis}
Y.~Kawai, {\em Development of a Hybrid Photon-Detector Module for Next
  Generation Water-Cherenkov Detectors}.
\newblock PhD thesis, The Graduate University for Advanced Studies (SOKENDAI),
  2007.

\bibitem{kume_1983}
H.~Kume et~al., {\it {20 INCH DIAMETER PHOTOMULTIPLIER}},  {\em Nucl. Instrum.
  Meth.} {\bf 205} (1983) 443--449.

\bibitem{Hamamatsu_R3899U}
{\it {{H}amamatsu {P}hotonics {K}.{K}., {R}3809{U}-61/-63/-64 data sheet}},
  {accessed July}, 2013.
\newblock
  {http://www.hamamatsu.com/resources/pdf/etd/R3809U-61-63-64\_TPMH1295E04.pdf}.

\bibitem{Blake}
A.~Blake, {\em WCSimAnalaysis Reconstruction Package}.
\newblock Cavendish Laboratory, University of Cambridge, UK.

\bibitem{Sanchez2012525}
M.~Sanchez and M.~Wetstein, {\it {Using Large Area Microchannel Plate
  Photodetectors in the Next Generation Water Cherenkov Neutrino Detectors}},
  {\em Nuclear Physics B - Proceedings Supplements} {\bf 229232} (2012) 525.
  Neutrino 2010.

\bibitem{SuperKalgo}
M.~Ishitsuka, {\em L/E Analysis of the Atmospheric Neutrino Data From
  Super-Kamiokande}.
\newblock PhD thesis, University of Tokyo, 2004.

\bibitem{Smy}
{M. Smy for the Super-Kamiokande Collaboration}, {\it {Low Energy Event
  Reconstruction and Selection in Super-Kamiokande-III}},  in {\em {Proceedings
  of the 30th International Cosmic Ray Conference}} ({R.~Caballero,
  J.C.~D'Olivo, G.~Medina-Tanco, L.~Nellen, F.A.~S\'{a}nchez,
  J.F.~Vald\'{e}s-Galicia}, ed.), vol.~5, pp.~1279--1282, {2008}.

\end{thebibliography}\endgroup

\end{document}